\documentclass[useAMS,usenatbib]{mn2e}   
\usepackage{graphicx}
 
\newcommand{\msun}{{\rm M}_{\solar}}
\newcommand{\Msun}{{\rm M}_{\solar}}
\newcommand{\solar}{\ifmmode_{\mathord\odot}\;\else$_{\mathord\odot}\;$\fi}
\newcommand{\kms}{\, \rm{km}\,  \rm{s}^{-1}}
\newcommand{\kpc}{\, \rm{kpc}}

\newcommand{\Aone}{{\it A}$_1$}
\newcommand{\Atwo}{{\it A}$_2$}
\newcommand{\Hone}{{\it H}$_1$}
\newcommand{\A}{{\it A}}
\newcommand{\B}{{\it B}}
\newcommand{\C}{{\it C}}
\newcommand{\R}{{\cal R}}
\def\ltsima{$\; \buildrel < \over \sim \;$}
\def\lsim{\lower.5ex\hbox{\ltsima}}

\begin{document}

\title{Secular bar formation in galaxies with significant amount of
	dark matter}
\author[Valenzuela \& Klypin]
       {Octavio Valenzuela and Anatoly Klypin \\
        New Mexico State University, Las Cruces, NM 88001 }
\date{Accepted ...;
      Received ...;
      in original form ...}
\maketitle

\begin{abstract} 
Using high resolution N-body simulations of stellar disks embedded in
cosmologically motivated dark matter halos, we study the evolution of
bars and the transfer of  angular momentum between halos
and  bars. We find that dynamical friction results in some
transfer of angular momentum to the halo, but the effect is much
smaller than previously found in low resolution simulations and is
incompatible with early analytical estimates. After 5~Gyrs of
evolution the stellar component loses only 5\% -- 7\% of its initial angular
momentum.  
 
Mass and force resolutions are crucial for the modeling of bar
dynamics. In low resolution (300 -- 500~pc) simulations we find that the
bar slows down and angular momentum is lost relatively fast. In
simulations with millions of particles reaching a resolution of
20-40~pc, the pattern speed may not change over billions of years. Our
high resolution models produce bars which are fast rotators, where the
ratio of the corotation radius to the bar major semi-axis lies in the
range $\R =1.2-1.7$, marginally compatible with observational
results. In contrast to many previous simulations, we find that bars
are relatively short. As in many observed cases, the bar major
semi-axis is close to the exponential length of the disk.

The transfer of angular momentum between inner and outer parts of the
disk plays a very important role in the secular evolution of the disk
and the bar. The bar formation increases the exponential length of the 
disk by a factor of 1.2 -1.5. The transfer substantially increases the
stellar mass in the center of the galaxy and decreases the dark
matter-to-baryons ratio. As the result, the central 2~kpc region is
always strongly dominated by the baryonic component. At intermediate
(3 -- 10~kpc) scales the disk is sub-dominant. These models demonstrate
that the efficiency of angular momentum transfer to the dark matter
has been greatly overestimated. More realistic models produce bar
structure in striking agreement with observational results.
\end{abstract} 


\section{Introduction} 
\label{sec:intro} 
Making predictions for the structure of individual galaxies in the
framework of cosmological models is a difficult task, that until
recently was mostly dominated by modeling of  dark matter 
dominated systems such as dwarf galaxies and low surface brightness
galaxies \citep[e.g.,][]{Moore:1994, FloresPrimack:94,
KravtsovCores:1998,BoschSwaters:2001,Lokas:01,deBlokMcGaughRubin:2001}. It
is generally accepted that models do not fare well on those tests
predicting too fast rotation in the central region. Excessive amount
of dark matter satellites is an additional complication for the
cosmological models
\citep{KlypinSatellites:99, MooreSatellites:99}.

Another layer of problems is on slightly larger scales: luminous parts  
of normal galaxies and masses in the range of $10^{10}-10^{12}\Msun$.  
The number of issues is quite large.  An excessive amount of  dark  
matter in the bulge may result in too few micro-lensing events  
\citep[e.g.,][]{ ZhaoMao:96, BinneyEvans:01}, but see also  
\citep{KlypinZS:02}. Mass modeling of the Milky Way galaxy on  
different scales is a complicated and quite uncertain issue  
\citep[e.g.,][]{DehnenBinney:98}. For  halos predicted by  
cosmological models the conclusions change from very pessimistic   
\citep{NavarroSteinmetz:00,HernandezAvila:01}  
 to neutral \citep{EkeNavarroSteinmetz:01} to more optimistic  
 \citep{KlypinZS:02}.  

One of the contentious issues is the pattern speed of bars. Using
analytical arguments and a combination of N-body simulations with
rotating solid bars \citet{Weinberg:85,HernquistWeinberg:92} argued
that a non-rotating dark matter will exert so much tidal friction that
a bar should lose its angular momentum in few rotation periods. For a
rigid rotating bar with mass comparable to the mass of the spheroid
(including dark matter) inside the bar radius
\citet{HernquistWeinberg:92} predict that the bar ``sheds all its
angular momentum typically in less than $\sim 10^9$~yr''.  If true,
that would make impossible for bars to exist in galaxies with dark
matter halos. There are no doubts that the dynamical friction operates
when a bar rotates inside a dark matter halo. Yet, the rate and the
amount of the angular momentum lost by the disk is still a matter of
debate. Fully self-consistent N-body simulations with live bars,
disks, and dark matter halos \citep{CombesSanders:81,
SellwoodAthanassoula:86, DB:99, DB:2000, AthanassoulaMisiriotis:02}
show that bars slow down far less than what has been predicted by
\citet{Weinberg:85,HernquistWeinberg:92}. Bars in simulations rotate
for billions of years, defying the theoretical expectations. Yet, bars
in numerical models are observed to slow down. \citet{DB:99,DB:2000}
find that in models, in which initially the dark matter constitutes
about the same amount of mass as the stellar disk in the central part
of the galaxy, the baryonic component (disk + bar) loses about 40
percent of it angular momentum over $\sim 10$~Gyrs. The bar pattern
speed declines by a factor of five after the bar forms. This produces
a bar which rotates more than twice slower than the observed bars. The
speed of bar rotation is often characterized by the ratio $\R$ of
radius of corotation to the large semi-axis of the bar.  


Pattern speeds for real galaxies can be estimated using the method of
\citet{TremainWeinberg:84}.  For a few galaxies -- most of them are
S0's -- the ratio was measured and is in the range $\R \approx
1.1-1.7$ \citep{Merrifield1995,Gerssen1999, Aguerri2003}.  Bars in
models of \citet{DB:99,DB:2000} have $\R =2.0 -2.6$, which contradicts
the observed values. 
   
A less known but not less important problem of bars is their
length.  \citet{CombesElmegreen:93} compare lengths of bars in
N-body simulations with those observed in galaxies.  Even though a
good agreement with observations was found, the models either did not
include a live halo or included only a bulge component and no halo at
all.  In more recent simulations  with  live halos (dark matter
dominated or not) the bar semi-major axes are 2-4 times longer than
the initial disk exponential length \citep{DB:99,DB:2000,
AthanassoulaMisiriotis:02}.  These long bars  contradict
observations because real galaxies typically have bar lengths 
0.5-1.5 of the present time exponential disk scale
\citep{Elmegreen:85}. For example, the bar in our Galaxy is 3-3.5~kpc
long \citep{BlitzSpergel:91,Zhao:96,Freudenreich:98, Gerhard:02} which
is close to the disk scale length of 2.5-3.5~kpc
\citep{DehnenBinney:98,Hammersley:99, Gerhard:02}. 
In the case of the
models studied by \citet{DB:99,DB:2000} the bar extended over the
whole disk. That would correspond to 15~kpc, if we scale the models to
match our Galaxy. Scaling the dark matter dominated model MH of
\citet{AthanassoulaMisiriotis:02} to our Galaxy, the radius of the bar
would be 14~kpc (about 4 times the initial exponential length).
 
The arguments -- the slowing down of bars and the lengths of bars --
against significant amount of the dark matter in the central parts of
normal galaxies have problems on their own. The arguments are based on
numerical simulations, which suffer from two types of problems:
numerical effects and initial conditions. Numerical resolution is very
limited in most simulations. It is often emphasized that models should
have a large number of particles to suppress the two-body scattering
and to reduce the noise \citep[e.g.,][]{Weinberg:98, DB:2000}. Yet,
the force resolution is another important effect, which is often
overlooked. For example, the formal (one cell) resolution in
simulations of \citet{DB:2000} was 1/5 of the disk scale length. The
scale hight was 1/2~cell, which means that vertical structure was not
resolved. If we scale the disk scale length to 3.5~kpc of our Galaxy,
then we get the formal resolution of 700~pc. To make the things worse,
the real resolution in the Particle Mesh code used by \citet{DB:2000}
is not the formal (cell) resolution, but about twice the formal
resolution.  For \citet{DB:99} and \citet{DB:2000} simulations this
corresponds to 1.4~kpc, which is clearly not enough to mimic the disk
of our Galaxy. The resolution in simulations of \citet{Fux:97} was
1.8\kpc\ at the solar distance.

There are number of reasons why high resolution (both mass and force)
is needed for simulations of the bar formation. (1)  {\it The central
region of galactic models.} Propagation of waves through the center
and swing amplification are prime suspects for bar instability
\citet{Toomre1981, BinneyTremaine1987}. Simulations must have
sufficient resolution to allow correct treatment of these waves. High
resolution at the center is also required because during the bar
formation the density at the center increases very dramatically
\citep[e.g.]{AthanassoulaMisiriotis:02}. This infall of the stellar
component to the center should be accurately tracked: the distribution
of mass in the central region may affect the structure of the bar
\citep{Norman1996,Berentzen1998,ShenSellwood:03}.  (2) {\it The
vertical structure of the disk.} Disks of galaxies are very thin -
only a couple hundred kpc.  Resolving the vertical structure of disks
is a challenge for numerical simulations: even optimistically the
resolution should be not worse than 100~kpc. Tracking the vertical
evolution of the disk is even more difficult than one naively
expects. At some stages of the evolution the disk develops waves which
oscillate in vertical direction
\citep{FriedmanPolyachenko1984,MerrittSellwood1994}. Instabilities
related with these waves (e.g. the fire-hose instability) are often
blamed for sudden thickening of the disk \citep{Raha1991,
AthanassoulaMisiriotis:02}. In this case the code should be able to
maintain both the vertical structure and to treat sufficiently
accurately the upward and downward
displacements of the thin disk. The vertical oscillations combined with random radial
stellar velocities produce coupling of the vertical
and the radial directions. As the result, high resolution in the
vertical direction must be accompanied with the same resolution in the
radial direction.  (3) {\it Non linear coupling of waves}. The role of
interactions between different modes is poorly understood
\citep{Contopoulos1981,FriedmanPolyachenko1984}. Theory operates with
linear perturbations and typically ignores the nonlinearities
\citep[e.g.][]{NelsonTremaine1999}. This does not mean that the
nonlinearities are not important: we simply do not know how to handle
them analytically.  There are no doubts that these effects exist: bars
have very large density contrasts. As such, they are nonlinear
features. Even more difficult problems are related with the reaction
of the dark matter to the small-scale effects in the disk. The dark
matter reacts to what happens in the disk sometimes even in a resonant 
fashion \citep{Athanassoula2003}. In turn, disk reacts to the
disturbances  in the halo. This non-linear interaction is clearly
important for the long-term evolution of the bar \citep{Athanassoula2003}.
Numerical simulations are the only available tools to study
the effects. Coupling of long and short waves (e.g. the short vertical
frequencies with long $m=2$ modes) require very high resolution. It
should be noted that imposing a high force resolution without adequate
mass resolution can be a disaster: simulations would amplify
small-scale shot noise potentially leading to incorrect predictions.

Initial setup is questionable in many cases when it comes to the
properties of the dark matter halo. If one wants to test predictions
of cosmological models, those predictions should be used for setting
initial conditions of numerical models.  For example, halos predicted
by cosmological models are very large and should extend to 200-300~kpc
for galaxies of the size of our Galaxy \citep{KlypinZS:02}. Halos in
simulations are much smaller than that. For example, \citet{Fux:97}
used halo truncated at 38~kpc. Models of \citet{DB:2000} were larger -
40~kpc, but still too small as compared to the virial radius of the
halo of our Galaxy. The density profile in the outer part of a halo
(radii larger 20~kpc for the Milky Way galaxy) should decline as
$\rho_{\rm dm}\propto r^{-3}$ \citep{NFW:97}. This is very different
from the profiles used in many simulations of bar formation. For
example, \citet{Fux:97} used exponential profile $\rho_{\rm DM}\propto
\exp(-r/9.1\kpc)$.  \citet{AthanassoulaMisiriotis:02} used density
profile $\rho\propto \exp(-r^2/(35\kpc)^2)/[r^2 +(1.75\kpc)^2]$
truncated at 52.5\kpc. \citet{DB:2000} used polytrops $n=3$, which
have little relation with the expected dark matter profile.

The large size and mass of cosmological dark matter halos affect the
motion of halo particles even in the central region. Because more
massive halos produce deeper potential wells and have larger escape
velocities, dark matter particles move faster in the central
region. In turn, larger velocities of particles make the halo more resistant to
interactions with the stellar disk. \citet{Athanassoula2003} argues
that ``hotter'' halos absorb less angular and result in slower
decline of the bar pattern speed.


Both problems of the existing numerical simulations -- the resolution
and the initial conditions -- motivate us to simulate with high
resolution models, which use more realistic initial conditions.  We use
the traditional approach to form a bar: bars are formed by dynamical
instability in initially featureless equilibrium disk rotating inside
halos \citep{Hohl:97,OstrikerPeebles:73,Miller:78}.  In
section~\ref{sec:models} we describe initial conditions for our models
and give details of the code used to run our simulations. The
evolution of models is discussed in section~\ref{sec:evolution}. In
section~\ref{sec:Angmom} we present results on the angular momentum of
different components. The pattern speed and the structure of bars are
presented in section~\ref{sec:BarStructure}. We discuss our results
and compare them with the previous results in 
section~\ref{sec:discussion}. A brief summary of our conclusions is
given in section~\ref{sec:conclusions}.

\section{Models}
\label{sec:models}
\subsection{Initial conditions: densities and velocities}
\label{sec:IC}

We study models, which initially have only two components: an
exponential disk and a dark matter halo. No initial bulge or  bar are
used. Initial conditions for the systems are generated using 
\citet{Hernquist:93} method. We use the following approximation for
the density of the stellar disk in cylindrical coordinates:
\begin{equation}
 \rho_d(R,z) =\frac{M_d}{4\pi z_0R_d^2}e^{-\frac{R}{R_d}}
                       sech^2(z/z_0),
\label{eq:expdisk}
\end{equation}
where $M_d$ is the mass of the disk, $R_d$ is the exponential length,
and $z_0$ is the scale height. The later was assumed to be constant
through the disk. The disk is truncated at five exponential lengths
and at three scale heights: $R<5R_d$, $z<3z_0$. Disk mass was slightly
corrected to include the effects of truncation. The vertical velocity
dispersion $\sigma_z$ is related to the surface stellar density
$\Sigma$ and the disk height $z_0$:
\begin{equation}
 \sigma^2_z(R) =\pi G z_0 \Sigma(R),
\label{eq:vertical}
\end{equation}
where $G$ is the gravitational constant. The radial velocity
dispersion $\sigma_R$ is also assumed to be directly related to the
surface density:
\begin{equation}
 \sigma^2_R(R) =A e^{-\frac{R}{R_d}}.
\label{eq:radial}
\end{equation}
The normalization constant $A$ in eq.~(\ref{eq:radial}) is fixed in
such a way that at some reference radius $R_{\rm ref}$ the radial
random velocities are $Q$ times the critical value needed to stabilize
a differentially rotating disk against local perturbations:
\begin{equation}
 \sigma_R(R_{\rm ref}) =Q\frac{3.36G\Sigma(R_{\rm ref})}{\kappa(R_{\rm
 ref})},
\label{eq:radiall}
\end{equation}
where $\kappa(R_{\rm ref})$ is the epicycle frequency at the reference
radius. The reference radius is chosen to be $R_{\rm ref}=5$~kpc for
our models. Results are very insensitive to the particular choice of
the reference radius because the stability parameter $Q$ does not
change much for radii from $R_d$ to few $R_d$. We also consider a
family of models where $Q$ is fixed along the disk.  In this case
eq.(\ref{eq:radiall}) gives the radial dispersion at every point. The
fast increase of the epicycle frequency $\kappa$ in the central region
overcomes the exponential growth of density making the disk colder in
the centre.

The tangential ($\phi$) component of the rotational velocity and its
dispersion are found using the asymmetric drift and the epicyclic
approximations:
\begin{eqnarray}
 V_\phi^2 &=& V_c^2 -\sigma_R^2\left(\frac{2R}{R_d}+
                  \frac{\kappa^2}{4\Omega^2}-1\right), \\
\sigma_\phi^2 &=& \sigma^2_R \frac{\kappa^2}{4\Omega^2}.  
\label{eq:phivelocity}
\end{eqnarray}
Here the circular velocity $V_c$ is found as the quadrature sum of the
halo and the disk contributions. For the disk component we use the thin-disk
approximation.

We assume that the dark matter density profile is described by the NFW profile
\citep{NFW:97}:
\begin{eqnarray}
 \rho_{\rm dm}(r) &=& \frac{\rho_s}{x(1+x)^2}, \ x=r/r_s, \\
 M_{\rm vir} &=& 4\pi\rho_s r^3\left[ \ln(1+C) -{C\over 1+C}\right],
   \ C={r_{\rm vir}\over r_s},
\label{eq:NFW}
\end{eqnarray} 
where $M_{\rm vir}$ and $C$ are the virial mass and the concentration
of the halo.  For given virial mass the virial radius of the halo is
found assuming a flat cosmological model with matter density parameter
$\Omega_0=0.3$ and the Hubble constant $H_0=70\kms {\rm Mpc}^{-1}$.

Knowing the mass distribution of the system $M(r)$, we find the radial
velocity dispersion of the dark matter:
\begin{equation}
 \sigma_{r, {\rm dm}}^2 ={1\over \rho_{\rm dm}}\int_r^{\infty}\rho_{\rm dm}\frac{GM(r)}{r^2}dr.
\label{eq:radialDM}.
\end{equation}
The other two components of the velocity dispersion are equal to
$\sigma_{r, {\rm dm}}^2$. In other words, the velocity distribution is
isotropic, which is a good approximation for the central parts of the
dark matter halos \citep[e.g.,][]{Colin:00}. Eq.(\ref{eq:radialDM})
ignores  non-spherical deviations of the gravitational acceleration
for the dark matter. At each radius we find the escape velocity and
remove particles moving faster than the escape velocity.

The dark matter halo is truncated at the virial radius, which for our
models is at 250~kpc -- 300~kpc radius. During the evolution of the
system, the truncation results in a gradual smearing of the outer
boundary of the halo. Some particles move to radii larger than the
virial radius. They are not replaced by other particles, which, in true
equilibrium, would come from the region outside the truncation
radius. This creates a rare-faction wave moving inside the halo. The
wave never reached the central part of the halo and was always outside
the central 100~kpc region.

The distributions eqs.~(\ref{eq:expdisk}-\ref{eq:radialDM}) are used
to produce initial coordinates and velocities of particles. The space
is divided into bins. We estimate the expected number of particles
inside each bin. We than randomly place specified number of
particles inside the bin. Spherical shells are used to initiate  dark
matter particles. For the disk we use cylindrical shells along radius and
equal-spaced bins along the axis of rotation.  Velocities are picked
from an appropriate Gaussian distribution truncated at the escape
velocity. The number of particles was typically 10-20 in each bin. The
resulting distribution of particles was tested against expected
analytical expressions.

The disks are realized with particles of an equal mass. The dark
matter halo is composed of particles of variable mass with small
particles  placed in the central region and larger particles at
larger distances.  We double the particle mass as we move further and
further from the centre. In total we use 5 mass species with the mass
range of 32.  The region covered by small mass particles is $\sim
40$~kpc -- much larger than the size of the disk. In order to reduce
the two-body scattering each dark matter particle in the central
region has the same mass as a disk particle.

This procedure is designed to reduce the number of particles and, at
the same time, to allow us to cover a very large volume. For example,
in  model \Aone\ we have 3.55 million particles inside $\sim 300$~kpc,
of which 2.4 million are small particles. We would need 9.5 million
particles, if particles of equal mass were used.  In the course of
evolution some of massive particles come closer to the centre, but
their number was always very small.  

\subsection{Choice of parameters}
\label{sec:parameters}

The approximation eq.~(\ref{eq:NFW}) formally is valid only for pure
dark matter halos without baryons. Contraction of baryons in the
process of formation of the disk should significantly
affect the dark matter. We try to mimic the effects of the contraction
by changing the halo concentration and by tuning parameters of the
halo and the disk in such a way that the distribution of the dark
matter roughly approximates more realistic
models. \citet{KlypinZS:02} presented such models for the Milky Way and for
the M31 galaxies. The models used realistic cosmological dark matter halos
and satisfied numerous observational constraints. To some degree we
mimic  two favored models in that paper: models $A_1$ and $B_1$, which
are described in detail in \citet{KlypinZS:02}.

The first model ($A_1$) was designed assuming that the angular
momentum of the dark matter was preserved during baryonic
collapse. The conservation of  angular momentum leads to a specific
prediction that in the central $\sim 5$~kpc the density of the dark
matter closely follows the density of the baryonic component. The
second model ($B_1$) includes effects of an exchange of  angular
momentum between the baryons and the dark matter. The dark matter 
gains angular momentum through the dynamical friction during early
stages of disk and bulge formation, when non-axisymmetric perturbations
are expected to be large. Somewhat similar scenario was proposed by
\citet{El-ZantShlossman:01}. Model $B_1$ 
also had significant amount of the dark matter in the central region,
but less than in model $A_1$. For example, the crossing point, at
which the contribution of the dark matter to the circular velocity is
equal to that of the baryons, is shifted from 5~kpc in model $A_1$
to 12~kpc in  model $B_1$.

In order to reproduce the effects of the contraction and the exchange
of the angular momentum, we increase the concentration of the dark matter
halos. Our models have $C=15$ as compared with $C=12$ for models of
\citet{KlypinZS:02}. We also adjust parameters of our models (virial
mass,  exponential length and mass of the disk) to roughly mimic 
important properties of \citet{KlypinZS:02} models in the central
20~kpc. For example, our models (\Aone\ and \Atwo) have a
sub-maximum disk (contributions of the disk and the halo to the
circular velocity are approximately equal) at radii $R<5$~kpc with
halo dominating at larger radii. The mass of the disk was $4.28\times
10^{10}\Msun$, which is close to the mass of disk+bulge in  model
$A_1$ of \citet{KlypinZS:02}. 

The only difference between our two models
\Aone\ and \Atwo\ is in the velocity dispersions of disk particles.
 For model \Aone\ we use eq.(\ref{eq:radial}) with the normalization
 provided by eq.(\ref{eq:radiall}) at the reference radius.  In the
 case of model \Atwo\ the eq.(\ref{eq:radiall}) is used at all radii
 (not just at one radius) and the eq.(\ref{eq:radial}) is not used at
 all.  In other words, model \Atwo\ has the {\it same} $Q$ parameter
 at all radii. The main difference between the models is that model
\Atwo\ is ``colder'' in the central region. In the
central $\sim 1$~kpc region model \Atwo\ has significantly
smaller radial and azimuthal velocity dispersions. Both models have
the same circular velocity and the same vertical velocity dispersion.

Our next model (\B) has a smaller halo and a more massive disk
($6\times 10^{10}\Msun$). The disk  mass is equal to  the sum of
the disk and the bulge masses in model $B_1$ of
\citet{KlypinZS:02}.  We have a third model called \C. 
This model has  a disk mass similar to that in models \A,
but both radial and vertical scale lengths are shorter. The halo
has the same mass as halo in model \B  but concentration is 
considerable larger.

\begin{figure} \includegraphics[width =0.47\textwidth]{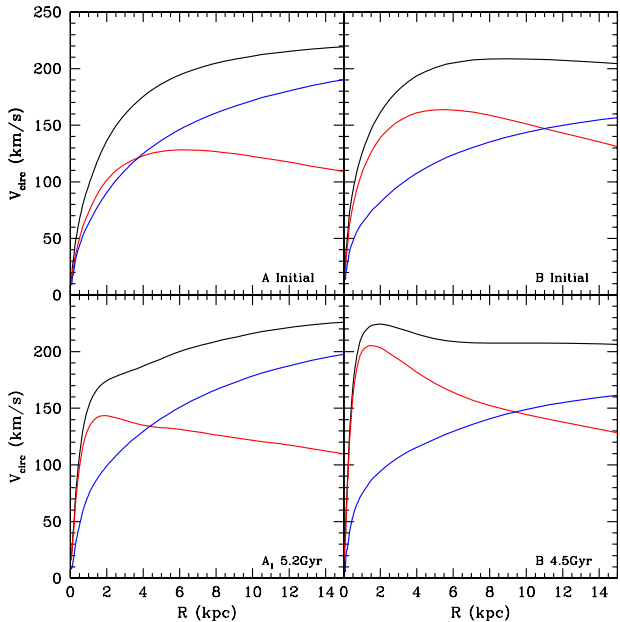}
    \caption{Initial and final circular velocities for model \Aone\
    (left panels) and for  model \B\ (right panels). Dashed curves show
   the contribution of the stellar component $\sqrt{GM(r)/r}$. The dot-dashed
   curves are for the dark matter. Final models show the total
   stellar contribution:  disk and  bar contributions are combined. Initially
   model \Aone\ has a sub-maximum disk with the density of the disk being
   close to the density of the dark matter inside the central
   3~kpc region. Model \B\ is more dominated by the disk component: the
   crossing point of the disk and dark matter is at 9~kpc. } 
  \label{fig:Vc} \end{figure}

Parameters of our models are presented in
 Table~\ref{table:initial}. Top panels in Figure~\ref{fig:Vc} shows
 the initial circular velocities $V_c =\sqrt{GM(r)/r}$ of the disk and
 the halo for the models.

   \begin{figure} \includegraphics[width =0.47\textwidth]{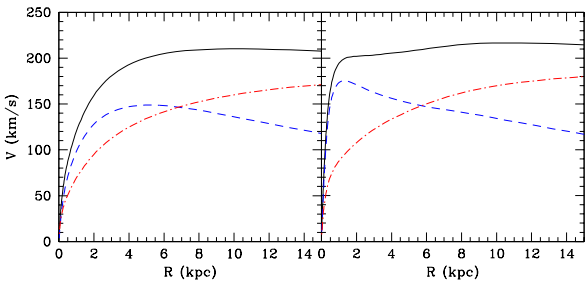}
   \caption{Initial (left panel) and final (right panel) circular
   velocities for model \C.  Dashed curves show the contribution of
   the stellar component $\sqrt{GM(r)/r}$. The full and the dot-dashed  
   curves are for the total circular velocity and for the contribution
   of the dark matter correspondingly.  } 
   \label{fig:VcLarge} \end{figure}

  \begin{figure} \includegraphics[width =0.47\textwidth]{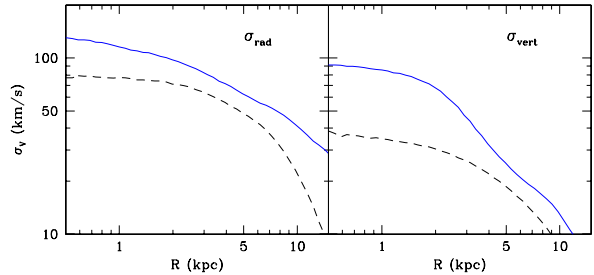} 
   \caption{Initial (dashed curves) and final (full curves) rms 
velocities of the stellar component for the model \C. The left panel shows the 
radial and the right panel shows the vertical ($z$) components of the  
velocity.  Initially the rms velocities are exponential with the ratio
of $\sigma_{\rm rad}/ \sigma_{\rm vert}\approx 2.5$. Formation of the
bar ``heats'' the vertical component more than the radial one. Outside 
the bar ($R> 5\kpc$) most of the change in the rms velocities occurs
in the first 1 Gyr when spiral arms form and dissipate. Notice that in this region 
the radial and azimutal components of velocities are preferentially heated by 
the spiral arms. } 
   \label{fig:sigLarge} \end{figure}

\begin{table} 
\caption{Initial parameters of models}
\label{param}
 \begin{center}       
 \begin{tabular}{|lllcccc} \hline     
\multicolumn{2}{l} {Parameter}  $A_{1}\&A_{2}$  &B   &C  \\\hline                                          
 Disk Mass ($10^{10} \Msun$)  & $4.28$  & $6.0$  & $4.8$ \\      
Total Mass ($10^{12} \Msun$) & $2.04$  & $1.0$  &$1.0$ \\  
Disk exponential length (kpc)      & 3.5  & 3.0  & 2.9 \\ 
Disk exponential height (kpc)      & 0.25  & 0.25 & 0.14  \\ 
Stability parameter $Q$  & 1.2 & 1.3  &1.2 \\ 
Halo concentration after  & 15 & 15 & 19 \\   
\quad baryonic contraction $C$ \\ 
Total number of particles ($10^5$)& $35.5$ & $7.8$  & $ 97.7$ \\    
Number of disk particles ($ 10^5$) & $2$ & $1.1$  & $ 12.9 $ \\    
Particle mass ($10^5 \Msun$) & $2.14$ & $5.45$ & $0.37$ \\   
Maximum resolution (pc) & 22 & 43.6 & 100.\\   
 \end{tabular} 
 \end{center}
\label{table:initial}
 \end{table}

There are important differences between our models and the models
described in \citet{KlypinZS:02}.  The later are realistic models,
which, in addition to the exponential disk and the halo, have also a nucleus
and a triaxial bar. The goal of our paper is to study how a bar
develops in an unstable disk. Thus, our models initially do
not have either a bar or a bulge: all the stellar mass is in the
exponential disk. The bar and the bulge are expected to develop  from the
disk.   Initially, our models mimic to some degree 
\citet{KlypinZS:02} models. For example, the exponential disk length 
is $3-3.5\kpc$.  Yet the subsequent evolution of the models was quite
significant. Thus, the final parameters a rather different
from what was assumed at the beginning. Our final models do not
provide a fit to the Milky Way galaxy. Nevertheless, the gross
properties (e.g., the maximum circular velocity and the stellar
surface density at 8.5~kpc) are not far from what is observed in our
Galaxy. This is why we scale  different physical
quantities to astronomical units (kpc, $\Msun$, and so on). 

\section{Numerical simulations} 
\label{sec:simulations}
\subsection{Code}    
\label{sec:code}   

We use the Adaptive-Refinement-Tree (ART) $N$-body code
\citep{kkk:97,kravtsov:99} to run the numerical simulations analyzed in 
this paper.  The code starts with a uniform grid, which covers the 
whole computational box. This grid defines the lowest (zeroth) level 
of resolution of the simulation.  The standard Particles-Mesh 
algorithm is used to compute the density and gravitational potential 
on the zeroth-level mesh with periodical boundary conditions.  The 
code then reaches high force resolution by refining all high density 
regions using an automated refinement algorithm.  The refinements are  
recursive. A refined region can also be refined. Each subsequent 
refinement level has half of the previous level's cell size. This creates 
a hierarchy of refinement meshes of different resolution, size, and 
geometry covering regions of interest. Because each individual cubic 
cell can be refined, the shape of the refinement mesh can be arbitrary 
and effectively match the geometry of the region of interest. This 
algorithm is well suited for simulations of a selected region within a 
large computational box, as in the simulations presented below. 

The criterion for refinement is the local density of particles. If the 
number of particles in a mesh cell (as estimated by the Cloud-In-Cell 
method) exceeds the level $n_{\rm thresh}$, the cell is split 
(``refined'') into 8 cells of the next refinement level.  The 
refinement threshold depends on the refinement level. The threshold    
for cell refinement was low on the zeroth level: $n_{\rm 
thresh}(0)=2$.  Thus, every zeroth-level cell containing two or more 
particles was refined.  The threshold was higher on deeper levels of 
refinement: $n_{\rm thresh}=3$ and $n_{\rm thresh}=4$ for the first 
level and higher levels, respectively.  

During the integration, spatial refinement is accompanied by temporal 
refinement.  Namely, each level of refinement is integrated with its 
own time step, which decreases by factor two with each refinement.  
This variable time stepping is very important for accuracy of the   
results.  As the force resolution increases, more steps are needed to    
integrate the trajectories accurately.

\begin{figure} \includegraphics[width =0.45\textwidth]{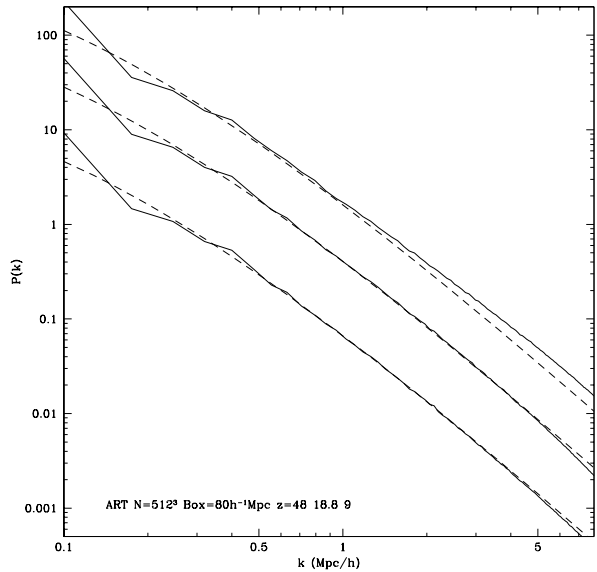}   	
\caption{ Test of the code: The growth of the power spectrum of
perturbations in the standard cosmological model with $512^3$
particles.  The dashed curves show the theoretical power spectra at
different moments. The full curves show results of the simulation.
The wiggles  of the simulated spectrum  at
small wavenumbers $k$ are due to the small number of long-wave
harmonics. They should be preserved (shifted upward).
 From bottom to top the curves are for redshifts $z=48,
18.8, 9$. At $z=9$ the fluctuations in the simulation have slightly
larger amplitude at large wavenumber due to nonlinear effects.  The
code accurately tracks the growth of about a million of independent
harmonics when the amplitude of fluctuations increases many times. }
 \label{fig:Power} \end{figure}

The  ART code has the ability to handle 
particles of different masses.  We use multiple masses 
to increase the mass (and correspondingly the force) 
resolution inside a region centered around the simulated galaxy.   

\subsection{Parameters of simulations}    
\label{sec:simlations}

The simulations presented here are run using $128^3$ zeroth-level grid
in a computational box of 2.86~Mpc for models \A and \B and 1.4~Mpc
for the model \C.  The models are placed in the center of the grid and
far from the periodic images. Because the size of the models are small
as compared with the size of the box, the effects of the periodical
images are very small and can be neglected. For example, at a distance
of 100 kpc from the center of the model \A the contribution of
periodic images is less that $7\times10^{-7}$ of the main galaxy
force. Even at 250~kpc, which is close to the virial radius for the
models, the tidal force is only $6\times 10^{-5}$ of the force from
the central image.  Bar formation and evolution are practically
unaffected by the periodic boundary conditions.

Model \Aone\ is simulated for
$8.2\times 10^9$~yrs with 550 zero-level time steps. The code reaches
10 refinement levels and has about 9 million cells. The number of time
steps at the highest resolution level is about 560,000, which gives     
the time step of $1.5\times 10^4$~yrs. With this time step it takes
3500 steps to make one orbital period at the 3~kpc radius. The formal
spatial resolution -- a cell size at the 10th level -- is 22~pc. At
about twice that value the force is close to the Newton's law. Central
6~kpc region was resolved with cells of the size not more than 100~pc.
Model \Atwo\ has similar parameters, but it was simulated for
7~Gyrs.

Model \B\ is simulated with a  smaller time step of
$1.02\times 10^4$~yrs (at the 9th level) for $4.5\times 10^9$~yrs. 
It reached the  highest resolution of 43.6~pc, but because of fewer
particles, the volume with high resolution is significantly smaller          
than in  models {\sl A}: there are only 3.9 million cells in model
\B. The number of steps at the highest 9th level is $4.4\times 10^5$.   

Model \C\ has the largest number of particles: almost 10 million of
which more than a million particles belong to the disk. The resolution
of this simulation was forced to be not better than 100~pc.  Because
of large number of particles, the whole disk (up to 14 kpc in radius)
is resolved with 100~pc cells.  The time step at that resolution was
$1.2\times 10^5$~yrs. The simulation box was 1.4~Mpc.  One of the
motivations for the model \C\ is to significantly reduce discreteness
effects. For example, our estimate of the two-body scattering time is
more than $10^5$~Gyrs for this model (see section\ref{sec:tbody}).

It is interesting to compare our numerical simulations with those  
presented in \citet{DB:99,DB:2000} and with more recent simulations of  
\citet{AthanassoulaMisiriotis:02}. It is straightforward to compare the 
resolution of  our simulations with that in \citet{DB:99,DB:2000}   
because we use the same Particle-Mesh algorithm. 
The formal (one cell) resolution in    
their simulations was 1/5 of the disk scale length. If we scale the   
length to 3.5~kpc, than we get the formal resolution of 700~pc, which 
is 30 times larger than in our  
models. \citet{AthanassoulaMisiriotis:02} used the TREE code with the   
Plummer softening of 0.0625 of the disk exponential length. Again,  
scaling it to 3.5~kpc, we get 220~pc. The force of gravity comes close   
to the Newton's law at approximately three Plummer softening lengths   
or at 660~pc. In our models \A\ this scale is 40-80~pc in the centre and  
is 200~pc for the entire 6~kpc region.     
\subsection{Tests of the code}    
\label{sec:codetests}   
 Extensive tests of the code and comparisons with other numerical
$N$-body codes can be found in \citet{kravtsov:99} and
\citet{knebe_etal:00}.  \citet{kravtsov:99} (Figure 6) shows the
accuracy of the gravitational force. The code was tested against known
analytical solutions: (1) the growth of small perturbations in the
expanding universe; (2) the spherical accretion (``the Bertschinger
solution''); (3) the collapse of a plane wave (the Zeldovich solution).  In
all tests the code performed extremely well. For example, in the case
of the spherical accretion the density at the center should increase
by 5 orders of magnitude over tested period of time. The code
reproduced that density increase as well as the whole radial density
profile.
Here we present results of two additional tests, which are more
relevant for the bar models: the growth rates of perturbations and the
stability of a self-gravitating equilibrium system.

Figure~\ref{fig:Power} shows the evolution of the power spectrum of
perturbations in the standard cosmological model. We use $512^3$
particles to set initial conditions at very high redshift $z=50$.  The
dashed curves in the Figure~\ref{fig:Power} show the theoretical power
spectra at different moments. The full curves show results of the
simulation.  At the initial moment the deviations of the simulated
spectrum from the analytical one at small wave numbers $k$ are due to a
small number of long-wave harmonics.  This is normal. The code
should preserve those deviations at later times, which it clearly
does.  This test shows that the code is able to accurately follow the
evolution of numerous waves over extended period of time during which
the waves increase their amplitude many times. For a code this is a
very difficult test. It probes every block of the code: gravity
solver, motion of particles, and density assignment.There are
numerous independent harmonics involved (millions in our case). Each
wave must grow independently on others in a well defined way. Because
the long waves have much larger amplitude, any erroneous small
coupling with short scales would ruin the power spectrum. The code was
able to accurately track the growth of all the waves during a period
when amplitude of the waves increases many times.

Tests also were performed to check the accuracy and stability of
bounded orbits of particles. For example, we set two particles in a
circular motion around each other and and run the code with the two
particles for hundreds of orbits. We do not find any drift of radius
of the orbit. Even a small systematic
loss or gain of angular momentum or any adverse effect when a particle
moves from one refinement level to another would be clearly detected
by the test. The code was tuned not to have any.

\begin{figure} \includegraphics[width =0.45\textwidth]{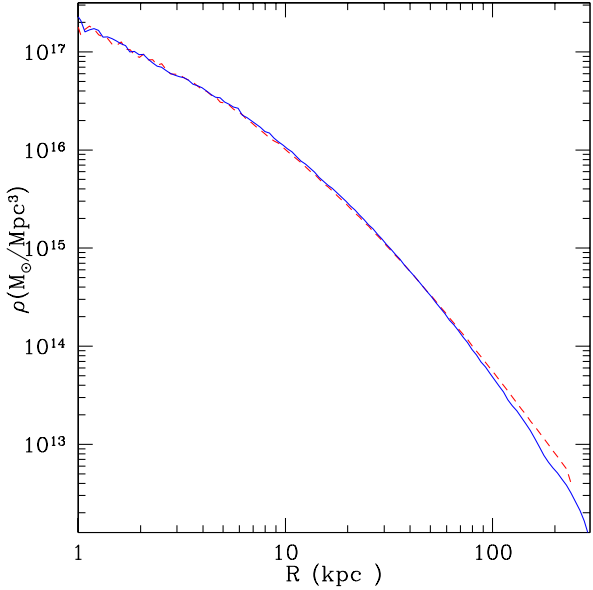} 	  
\caption{Test of the code: stability of an equilibrium NFW halo
 modeled with 200,000 particles for 5~Gyrs.  The dashed and the full
 curves show the initial and the final profiles.  The central part of
 the halo is potentially susceptible to different numerical effects
 and may be affected by the two-body scattering. The profile inside
 100~kpc shows no evolution.  The slight decline of density outside
 100~kpc is due to initial truncation of the halo at virial radius of
 250~kpc. }
 \label{fig:Pevol} \end{figure}

We present results of a test of this kind: a long-term stability of an
equilibrium self-gravitating configuration.  We use 200,000 particles
to set an equilibrium halo with the NFW density profile.  Three mass
species of particles are used, with the second specie being twice as
massive as the first one and the third one being four times as massive
as the first specie. The halo was truncated at the virial radius of
250~kpc.  Parameters of the halo simulation are presented in
Table~\ref{table:tbody}. The simulation is then run for 5~Gyrs. There
are different goals for this test. Effects of the two-body scattering
are discussed in the next section. Here we present the evolution of
the density profile. Figure~\ref{fig:Pevol} shows the initial and
final profiles of the halo. The only observed change in the profile is
the decline of the density close to the virial radius. This is caused
by the truncation of the configuration at the virial radius. A
rarefaction wave gradually moves in to inner parts of the halo. Yet the
effect of the wave is very small even at large radii. For example, the
density at 100~kpc decline only by 10 percent.  The central part ($r<
60\kpc$) of the halo does not show any measurable change of the
density.  Note that the crossing time for the central 2~kpc region is
only $10^7$yrs for typical velocity 200~$\kms$. Thus, the central part
was run for 500 crossing times. The NFW profile is a difficult test
because of the large densities and low velocity dispersion in the
central region. This cold and dense center potentially is unstable:
any numerical errors have a tendency to heat up the center and to
reduce its density. The code passes this tough test.

\begin{table} 
\caption{Parameters for the test Halo Simulation}  
\label{paramSim} 
 \begin{center} 
 \begin{tabular}{|lc} \hline 
Parameter \ Model       &   \Hone\    \\\hline  
Halo Mass ($\Msun$)  & $1\times 10^{12}$ \\ 
Halo concentration   & 15 \\ 
Total number of particles & $2\times 10^5$  \\      
Total number of mass species & $3$ \\   
Formal maximum resolution (pc) & 174 \\  
\end{tabular} 
\end{center} 
\label{table:tbody}  
\end{table}  

\subsection{Two-Body Scattering}    
\label{sec:tbody}   

Two-body scattering should be very small for systems, which we try to
model.  Given that our halo and disk are made of particles, every
precaution should be made to insure that discreteness does not affect
the evolution of the systems. On small scales this is done by imposing
a low limit on the number of particles inside a resolution element.
The force of gravity becomes Newtonian at about 2 cells. The code is
designed to have at least 4 particles in a cells for models \A~ and \B~
and about 20 for the model \C. Thus, inside one resolution element --
a sphere with a radius of 2 cells -- there are 130 and 670 particles for
models \A,\B~ and \C~ correspondingly. This large number of particles
means that the force acting on a particle can never be dominated by its
few closest neighbors. In other words, the code totally eliminates the
most damaging close collisions. They cannot possibly happen in the code we  use.

 Large distance encounters are possible and really happen in our models,
but their effect is small.  We use the simulation \Hone~ to
investigate the role of the two-body scattering in our numerical
experiments and, specifically, to study whether or not angular momentum
is affected by this process.  The angular momentum for each individual
particle should be preserved, if the model is truly
collision-less. The change of the orbital angular momentum of any given $i-$th
particle, $\Delta L_{i}= \vert \bf L_i - \bf L_{i,\rm init}\vert $,
could be attributed to numerical effects or to the two-body scattering
or to a combination of both. If it is dominated by the two-body
scattering, the random walk approximation is a reasonable model for
the change of orbital angular momentum (Chandrasekhar 1942). In this
case the rms value of the change of the orbital angular momentum
$\sqrt{ \Sigma \Delta L_{i}^{2}/N}$ should grow as the square root of
time.  The relaxation time associated with the two-body scattering can
be defined as the time when the rms change in the orbital angular
momentum is comparable to the rms value of the orbital angular
momentum $\sqrt{ \Sigma L_{i}^{2}/N }$.  This could be expressed as:
  
\begin{equation}  
\frac{\Sigma \Delta L_i^{2}}{\Sigma L_i^{2}}  = \frac{t}{T_{Relax}}                    
\label{eq:Relax} 
\end{equation}   

\begin{figure} \includegraphics[width =0.45\textwidth]{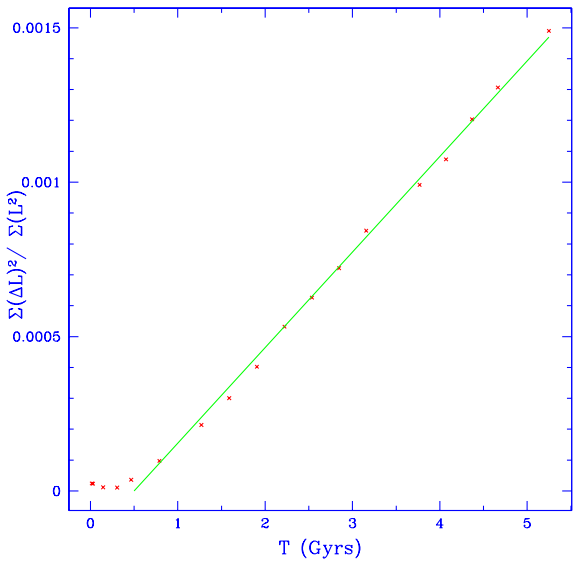}   
\caption{The test of the code for preserving the angular momentum of
 individual particles. The plot shows the time evolution of the average deviation of
 the angular momentum  $\Delta L_i= \vert \bf
 L_i - \bf L_{i,\rm init}\vert $ in the $H_1$ simulation with $2\times
 10^5$ particles (crosses).   The
 line is the $\Delta L \propto \sqrt T$ fit for the points after
 0.5~Gyro.  The slope implies the two-body relaxation time $T_{\rm
 Relax}= 3.2\times 10^3$~Gyrs, in agreement with the Chandrasekhar
 formula evaluated at 30~kpc.}
\label{fig:Scat} \end{figure}

The right hand size of eq.(\ref{eq:Relax}) is a linear function of
$t$ with the slope equal to the inverse of relaxation time
($T_{\rm Relax}$).  Figure~\ref{fig:Scat} shows the behavior of $ \Sigma
\Delta L^{2} /\Sigma (L_{i}^{2})$ over 5 Gyrs.  The linear behavior
shown in the plot supports the  random walk approximation
expressed in eq.(\ref{eq:Relax}).  The slope corresponds to a
relaxation time of $ 3.2 \times 10^{3}$ Gyrs.  As a test we can
compare our result with theoretical estimates.  Chandrasekhar (1942,
equation 2.380) gives an analytical expression for $T_{\rm Relax}$ due
to two-body scattering:

\begin{equation}    
 T_{\rm Relax} = 1.12\times 10^4 \times \frac{(\sigma_{3D}/20)^3}
    {n_1  m_1 m_2 \log_{10}\Lambda} {\rm Gyrs},
\label{eq:Trelax}  
\end{equation}  
where $m_1$ and $ m_2$ are the mass of the colliding particles in
solar masses, $\sigma_{3D}$ is the three dimensional rms velocity of
the scattered particles in $\kms$, $n_1$ the number density of the
scattering particles in $pc^{-3}$ and $ \ln\Lambda$ is the Coulomb
logarithm. An estimate of $T_{\rm Relax}$ based on a Foker-Planck
orbital average, which also uses velocity deviations instead of
energies, predicts a coefficient for eq.(\ref{eq:Trelax}) that differs
only by a factor of 0.6. We neglect this small correction. In the
case of scattering between particles of the first mass specie (the
dominant term) we should use in eq.(\ref{eq:Trelax}) the mass and the
number density of small particles.  For the Coulomb logarithm one
should take $ \ln\Lambda = \ln(b_{\rm max}/b_{\rm min})$, where $
b_{\rm max}$ and $ b_{\rm min}$ are the maximum and minimum values for
the collision impact parameter. A reasonable approximation for these
values is: $ b_{\rm max} = R_{\rm vir}= 252$~kpc and $b_{\rm min}=
{\rm force~ resolution} = 0.174$~kpc.  This gives $\ln\Lambda = 7.3$.
We calculate the value of the relaxation time using
eq.(\ref{eq:Trelax}).  If we take $ r=40$~kpc ( 25 \% mass radius),
$n=1.5 \times 10^{-10} pc^{-3}$, $m_{1}=2 \times 10^{6} ~\msun$,
$\sigma_{3D} = 171$~km/s, we obtain $T_{\rm Relax} = 3.6 \times
10^{3}$~Gyrs, which is consistent with the estimate obtained from our
halo simulation.

A disk-halo system modeled with particles potentially may experience a
large two-body scattering if there is a large difference in masses
between particles. This happened in old simulations, when massive
halos where modeled with relatively few particles to reduce the
computational time.  Our main simulations \A-\C~ use 5 mass species,
each mass specie being twice massive as the previous one. Thus, the
mass ratio of the most massive to the lightest particle is large and 
equal 16.  In the initial configuration the central 40~kpc of the halo
and of the disk is sampled only with the particles of the same (small) mass. This setup
is designed to minimize the artificial scattering. However, few
particles of the 5th (most massive) specie eventually reach the
central region.  How does this affect the two-body scattering
relaxation time estimate? An estimate of the relaxation time $T_{\rm
big}$ for particles of the 5-th specie giving energy to (``heating'')
the particles of the 1-st mass specie (small particles) can be
calculated using eq.(\ref{eq:Trelax}) by replacing the terms $n_1 m_1
m_2$ with $n_b m_b m_s$, where indexes $b$ and $s$ refer to big and
small particles correspondingly.  Thus, the relaxation time of the 1st specie
due to scattering with big particles scales as 
\begin{equation}
T_{b}   = \frac{ n_s m_s} {n_b m_b } T_{\rm Relax}  
\label{eq:B-S_scale}   
\end{equation} 
Even at late moments of evolution the number density of massive
particles in the central 1~kpc is 4 orders of magnitude smaller than
the number density of the lightest particles. Using those numbers we
obtain that $T_{b}$ is 625 times longer than the relaxation time for
the first specie $T_{\rm Relax}$.  There is an additional effect,
which increases this estimate even more.  Eq.~(\ref{eq:B-S_scale})
does not take into account the fact that large particles move a factor
of two faster when they move through the central region as compared
with the small particles. This happens because large particles have
large energies: initially they are placed at large radii. 
 In other words the two-body relaxation due
collisions with big particles is negligible in our simulations.

As an additional test for the effects of the two-body scattering we
study how the halo density profile evolves during the simulation:
two-body scattering should lead to a gradual decline in the central
density for the NFW halo.  Figure~\ref{fig:Pevol} shows the evolution
of the density profile of the halo simulation after 5~Gyrs.  The
central region of the halo has a higher density and a shorter
dynamical time.
However we do not find any change in the density profile even after 5
Gyrs. This confirms our estimates of the relaxation time-scale.

The simulation $H_1$ is used just as a test for code stability and for
estimates of the two-body scattering.  Our main simulations have many
more particles, and, thus the scattering should be even smaller for
them.  Using eq.~(\ref{eq:Trelax}) and assuming  virial equilibrium we
obtain a scaling relation: $T_{\rm 
Relax} \propto N/\ln N$ (Binney \& Tremaine 1987).
This relation allows us to scale-up the estimates for the relaxation time  based on
simulation $H_1$ to our simulations with a larger number of particles. 
For our disk-halo simulations with $3.5 \times 10^{6}$ and $10^{7}$
particles we get $T_{\rm Relax} =$ $ 4.5 \times 10^{4}$ Gyrs and $1.2
\times 10^{5}$ Gyrs respectively.  We can safely ignore any effects of
scattering in our disk-halo simulations.

\subsection{Finding the bar}
\label{sec:bar}
We follow \citet{DB:2000} and \cite{AthanassoulaMisiriotis:02} when
finding bars in our simulations.  We start with finding the centre of
the system. This can be done in different ways. \citet{DB:2000}
minimized the sum of distances of all particles relative to a
centre. We used a more simple algorithm: the centre of mass of stellar 
particles.  We then bin stellar particles using cylindrical shells
spaced equally in logarithm of the distance.  For each bin we find the
Fourier components of second harmonic of the angular distribution of
particles:
\begin{equation}
	a_2 = {1 \over N}\sum^N_{i=1} sin(2\phi_i), \quad 
	b_2 = {1 \over N}\sum^N_{i=1} cos(2\phi_i),
\label{eq:fourier}
\end{equation}
where $N$ is the number of stellar particles in the bin and $\phi_i$
is the angle of the $i-$th particles.  We then find the phase and the
amplitude of the second harmonic $A_2^2=(a_2^2+b_2^2)/2$.

Except for the very early stages of evolution, when the bar was just
emerging, it is easy to identify the bar. In the central region the
amplitude of the second harmonic $A_2$ changes gradually from bin to
bin and its phase $\phi$ is almost constant. At larger distances
(typically larger than 5~kpc) the amplitude declines and the phase
shows large variations. In the very centre the shot noise complicates 
the results.  Because of these considerations, we use bins with radii
in the range 0.5--3~kpc to find the amplitude and the phase of the
bars.  We find the bar amplitude and its phase as the
amplitude-weighted means at different radii: $\langle A_2\rangle =\sqrt{\langle
A_2^2\rangle}$, $\langle \phi
\rangle = \langle \phi A_2 \rangle /\langle A_2 \rangle$.  When
finding the average amplitude and the phase, we exclude bins with
$A_2$ less than 1/2 of the maximum amplitude found in the range of the
radii. Once the phase of the bar is found at different moments of
time, we estimate the frequency of bar rotation (the pattern speed) by
numerical differentiation: $\Omega_p =d\phi/dt$.

  \begin{figure*}
   {\includegraphics[width=\linewidth]{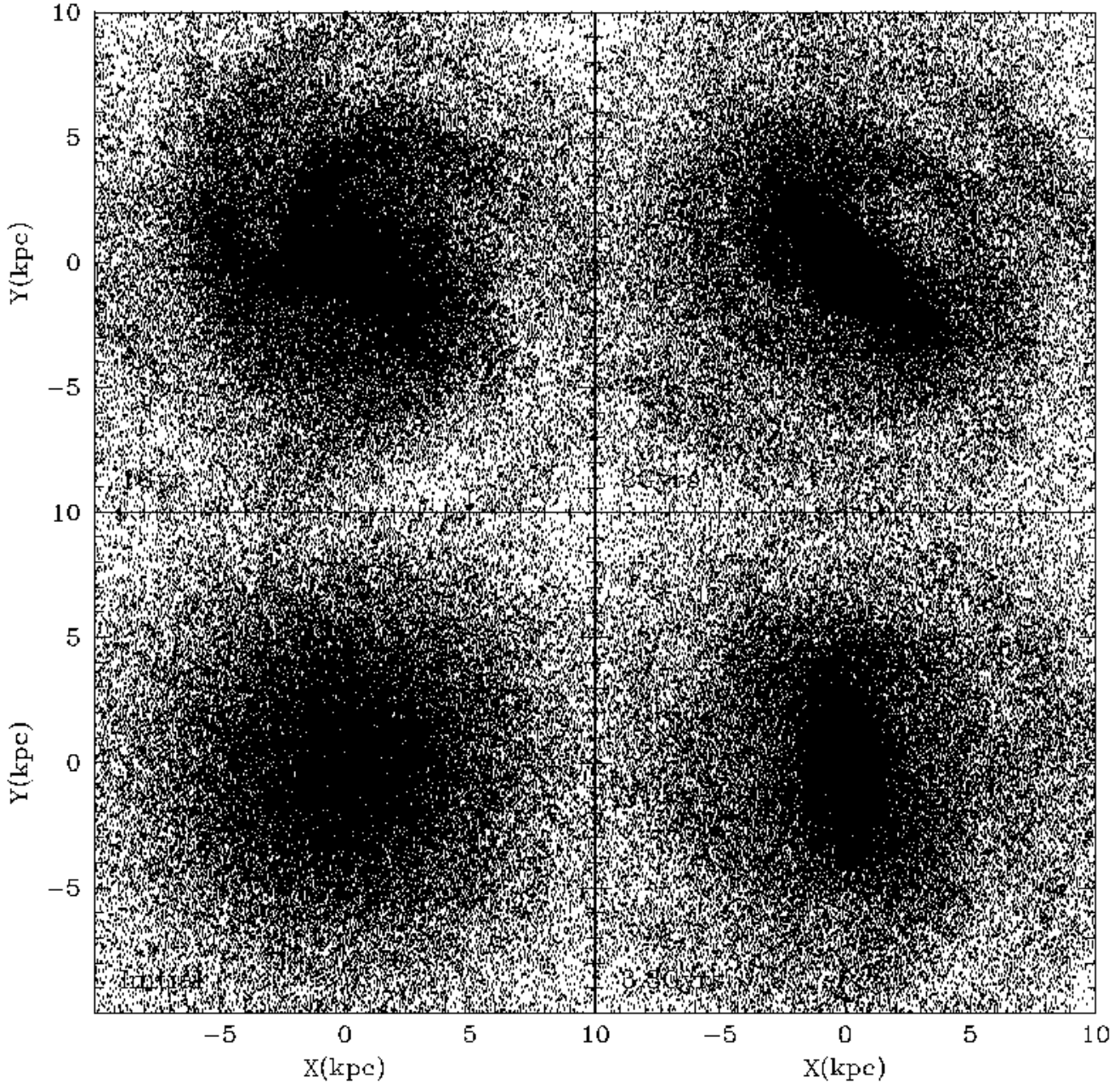}}
   \caption{Distribution of the stellar component at different stages
   of evolution for model \Aone. Starting from the left bottom 
   panel in the clock-wise direction the panels correspond to the initial
   moment; 1~Gyr; 2~Gyr; 3.3~Gyr. The disk rotates
   counter-clockwise. To avoid crowding we show only every fifth
   particle.  At the distance 5~kpc the disk made 20 orbital periods
   during 3.3~Gyr.} \label{fig:evol}
\end{figure*}

Finding the radius of the bar is a more difficult problem because bars
do not have clear boundaries. Discussion of different methods of
finding the bar major axis can be found in \citet{DB:2000} and in
\citet{AthanassoulaMisiriotis:02}. We use two different methods, each being
sensible and producing reasonable, but slightly different results.  We
find the surface density along and perpendicular to the bar. We then
define the radius of the bar as the radius at which the surface
density along the bar is twice of the surface density at the same
distance perpendicular to the bar. In the second prescription the bar
radius is defined as the radius at which the contours of the surface
density have axial ratio of two.  We also tried other prescriptions,
which are to some degree reminiscent to what was used by
\citet{DB:2000}. We find the major semi-axis as the radius at which
either the phase of the bar starts to deviate by more than 10 degrees
or as the radius at which the amplitude falls below 1/2 of its maximum
value. We find that typically the prescriptions based on the surface
density give the same results as those based on the amplitude and the
phase of the bar. For example, for  model \Atwo\ at
$t=6.2$~Gyrs the bar radii are: $R_{\rm bar}= 4.3\kpc$ ($A_{2, {\rm
max}}/2$), $=5.5\kpc$ ($\Delta\phi =10^{\rm o}$), $=4.2\kpc$
($\Sigma_{\rm major}(r_{\rm major})=2\Sigma_{\rm minor}(r_{\rm
major})$), $=5.3\kpc$ ($\Sigma_{\rm major}(r_{\rm major})=\Sigma_{\rm
minor}(r_{\rm major}/2)$). Here $\Sigma_{\rm major}(r)$ and
$\Sigma_{\rm minor}(r)$ are the stellar surface densities along the
major and the minor axes respectively. Nevertheless, we find that the
surface densities give more stable results because occasionally spiral
waves in the outer disk align with the bar and the amplitude/phase
conditions produce spurious results. 

We present the total range of the bar sizes produced by the two
methods based on the surface density profiles. The estimates vary by
20--30 percent.  We would like to emphasize that every estimate was
reasonable: the estimates in the middle of the range are no better
than the extremes.

\section{Evolution}
\label{sec:evolution}
Figure~\ref{fig:evol} shows the evolution of the stellar component in
model \Aone. Because the system is relatively hot and the dark matter
is significant in the central region, the bar develops slowly. After 
1~Gyr  (top left panel) the bar has not yet
appeared. Nevertheless, extended spiral arms are clearly present in
the disk. At that time the system has already made many rotations: at
the distance of 5~kpc one orbital period is equal to $T\approx
1.6\times 10^8$~yr. The bar develops after 2~Gyrs (top right
panel). At that moment its radius is $\approx 4-5$~kpc. The spiral
arms can still be seen, but they start to get weaker after that. At
the last shown moment (3.3~Gyr) only fragments of spiral arms can be
detected. The bar is still very strong, but its amplitude is visibly
smaller than at the peak at 1.8~Gyr. The bar gradually grows with
time. At later moments (5 -- 8~Gyrs) the major semi-axis of the bar is
$\approx 6\kpc$.  The evolution of model \B\ is qualitatively the
same, but it's initial evolution is more violent and is faster than in
the case of model \Aone.  

The evolution of model \Atwo\ remarkably differs from that of model
\Aone\ in spite of the fact that their initial parameters are almost the
same. The only difference between the models is in the initial random
stellar velocities: the central part of the disk is colder in model
\Atwo. The initial evolution is faster in model \Atwo. After 1~Gyr the
bar is already in place. The bar is shorter (1 -- 1.5~kpc) and
it rotates almost twice faster then the bar in model \Aone.  The bar
very slowly increases its length.  This slow evolution proceeds for
2~Gyrs.  At 3~Gyrs from the beginning the bar suddenly starts to grow
in length and its amplitude also increased significantly.

\begin{figure} 
	\includegraphics[width =0.5\textwidth]{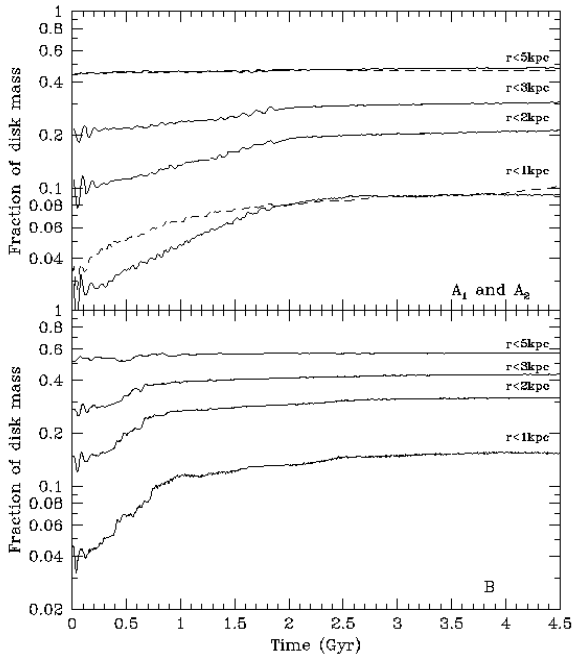}
	\caption{Evolution of  fraction of disk mass inside given
	radius. The top panel is for  models \Aone\ (full curves) and  \Atwo.
                The bottom panel shows model \B.
                The total mass inside 5~kpc practically does not
	change with time. Most of the mass exchange happens inside the   
	central 1 -- 2~kpc. The disk mass inside the central 1~kpc region
	increased by a factor 4 by the end of evolution. The
	time-scale of  evolution is about twice shorter in the case of
	the more disk-dominated model \B\ and the colder model \Atwo.}
\label{fig:Massevol} 
\end{figure}

Formation of bars is accompanied by significant redistribution of the
stellar mass.  Even a visual comparison of the bottom panels in
Figure~\ref{fig:evol} shows that the central surface density
increases. Figure~\ref{fig:Massevol} shows the evolution of the
fraction of stellar mass inside given radius. The total mass inside
5~kpc practically does not change with time. All the mass exchange
happens inside the central 1-2~kpc  and it  is very violent. For
example, the stellar mass inside the central 1~kpc region increases by
a factor 4 (5.5) after 5~Gyrs (8~Gyrs) of evolution. Note that the
time-scale of the initial evolution is about twice shorter in the
case of the more disk-dominated model \B\ and for the colder model
\Atwo.  

\begin{figure}
	\includegraphics[width =0.45\textwidth]{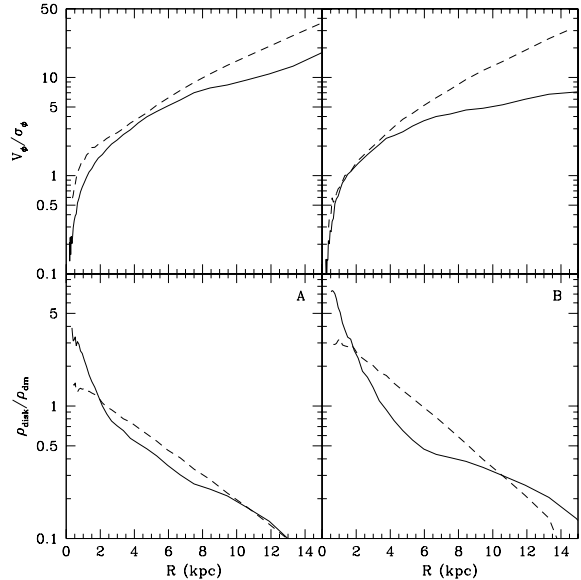}
	\caption{The baryon-to-dark matter ratios (bottom panels) and
	the ratio of the rotational velocity to the velocity dispersion
	(top panels) for model \Aone\ (left panels) and  model \B\
	(right panels). The dash curves show the initial models. The
	final models are shown with full curves. The central growth of
	the density goes at the expense of the stellar density in the
	middle 2 -- 8~kpc region. At those radii the final
	baryon-to-dark matter ratio went down producing dark-matter
	dominated disk. The top panels show that the heating of the
	disk preferentially happens in the outer regions outside of
	the bar. Spiral waves may cause the heating. }
\label{fig:profile} 
\end{figure}

Bottom panels in Figure~\ref{fig:profile} present more detailed
information on the change in the densities. In the central 2~kpc the
ratio of the stellar density to the dark matter density have increased
by a factor of two by the end of the evolution. As the result, even
model \Aone\ became significantly baryon-dominated in the centre,
the same effect was already observed in simulations by \citet{AthanassoulaMisiriotis:02}.
The Figure also shows that the central growth of the density goes at
the expense of the stellar density in the middle part ( 2--8~kpc ). At
those radii the final baryon-to-dark matter ratio was a bit
surprising: it went down producing a dark-matter dominated disk. This
effect is more pronounced in model \B, which was initially more
baryon-dominated than models {\sl A}. 

The dramatic changes of the disk component hardly affect the dark
matter: it stays almost unchanged. There is a 20 (40) per cent
increase of the dark matter density in the central 2~kpc region after
5 (8)~Gyrs of evolution . The rest of the halo is practically
unaffected in agreement with \citet{AthanassoulaMisiriotis:02}. 

Because of the violent processes in the central region, one would
naively expect to find a very strong heating of the disk in the
central part and much less in the outer parts.  The ratio of the
rotational velocity $V_{\phi}$ to the velocity dispersion
$\sigma_{\phi}$ characterizes the ``hotness'' of disk. The top plots
in the Figure~\ref{fig:profile} show just an opposite trend: the
heating is relatively stronger in the outer part of the disk. A possible
explanation for this is the heating by spiral waves that appeared in
the outer parts of the disk. Large amplitude spiral waves produce
non-circular motions and increase the azimuthal dispersion. After the
waves die out, they leave behind a hotter disk.

\section{Angular momentum}
\label{sec:Angmom}
Formation of a bar is accompanied and even may be driven by the
exchange of  angular momentum between different components of the
system. The exchange of  angular momentum between the quickly
rotating stellar component and the non-rotating dark matter is of
special interest because  is often used as one of the arguments
against the presence of a significant amount of the dark matter in the
central parts of galaxies.  Numerically it is a challenge to simulate
the two-component system: any artificial numerical coupling will
result in a transfer of some angular momentum from the stellar to the
dark matter component. For an axisymmetric system no transfer should 
happen. Once the bar or any other non-axisymmetric perturbation is formed,
the dark matter should react to it by generating a wake behind the
perturbation, which must result in a transfer of the angular momentum
to the dark matter. The exchange of the angular momentum is not
limited by the disk-dark matter interaction. The stellar component
also experiences the exchange between different parts of the disk,
which we find to be much more important for the evolution and for the
structure of bars.

\begin{figure*} 
	\includegraphics[width =\linewidth]{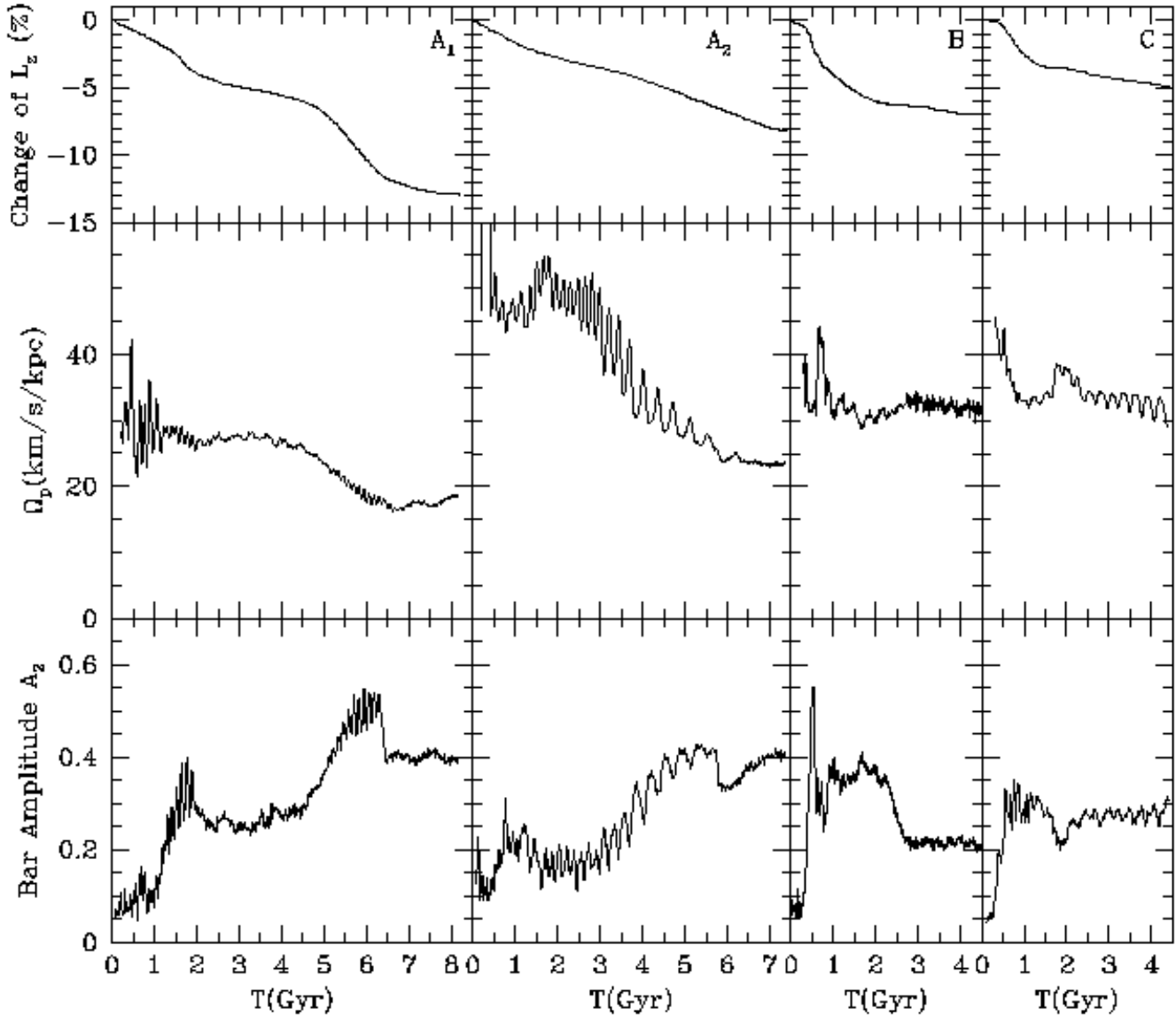}
	\caption{Evolution of the bar amplitude (bottom panels), the
	pattern speed (middle panels), and disk angular momentum for
	the models \Aone\ (left column of panels), \Atwo\ (middle
	column of panels) and \B\ (right column).  In all cases the
	angular momentum declines due to the dynamical friction with
	the dark matter. The decline is very small - only few percent
	for $\sim 5$~Gyrs of evolution. Note the complex behavior of
	bar amplitude and its correlation with the pattern speed. The
	pattern speed remains almost constant during periods, when the
	bar amplitude does not evolve (e.g., $t=2-5$~Gyrs for model
	\Aone or $t=5-7$~Gyrs for model \Atwo). Periods of fast
	increase of the bar amplitude correlate with the decline of
	the pattern speed (e.g.,$t=3-5$~Gyrs for model \Atwo). }
	\label{fig:Levol}
\end{figure*}

\begin{figure} 
	\includegraphics[width =0.5\textwidth]{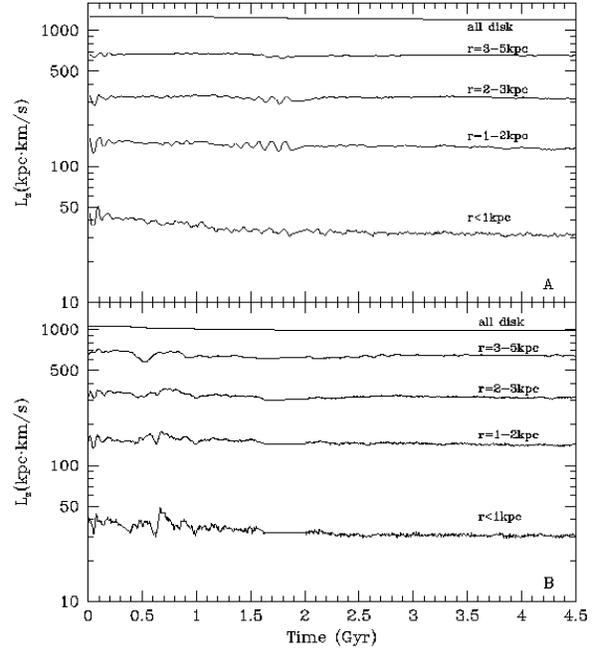}
	\caption{Evolution of the specific angular momentum of the
	disk for the \Aone\ (top panel) and the \B\ (bottom panel)
	models.  The top curves in each panel show that the
	total angular momentum of  disk particles declines very
	little -- few per cent after 4.5~Gyrs of evolution. The other
	curves present an average specific angular momentum for disk
	particles inside  spherical shells indicated in the
	plot. Only the very central region (radius less than 1~kpc)
	exhibits some decrease (a factor of 1.5) of  angular
	momentum. Outside the very centre the specific angular
	momentum does not change. } \label{fig:AngProfile}
\end{figure}

Figure~\ref{fig:Levol} presents the evolution of the z-component of
the angular momentum of the stellar component (disk + bar; top
panels), the pattern speed of bars (middle panels), and the bar
amplitudes (bottom panels) for all the models.
Figure~\ref{fig:AngProfile} shows details of the evolution of the
angular momentum for \Aone\ and \B\ models.  The total stellar angular
momentum clearly is declining with time, but the rate of the decline
is very small. After 5~Gyrs the change was $\approx 5$\% for all the
models. The disk of the model \Aone\ lost 13 per cent of the angular
momentum after 8~Gyrs of evolution. This level of the angular momentum
loss is at odds with the results of \citet{DB:2000}, who find about
40\% decline.  A large fraction of the angular momentum loss in the
model \Aone\ happened during a short period (5-6~Gyrs after the
start), when the bar was increasing its amplitude. When the growth
abruptly stopped at $t\approx 6$~Gyrs, the rate of the decline also
slowed down dramatically.

\begin{figure} 
	\includegraphics[width =0.5\textwidth]{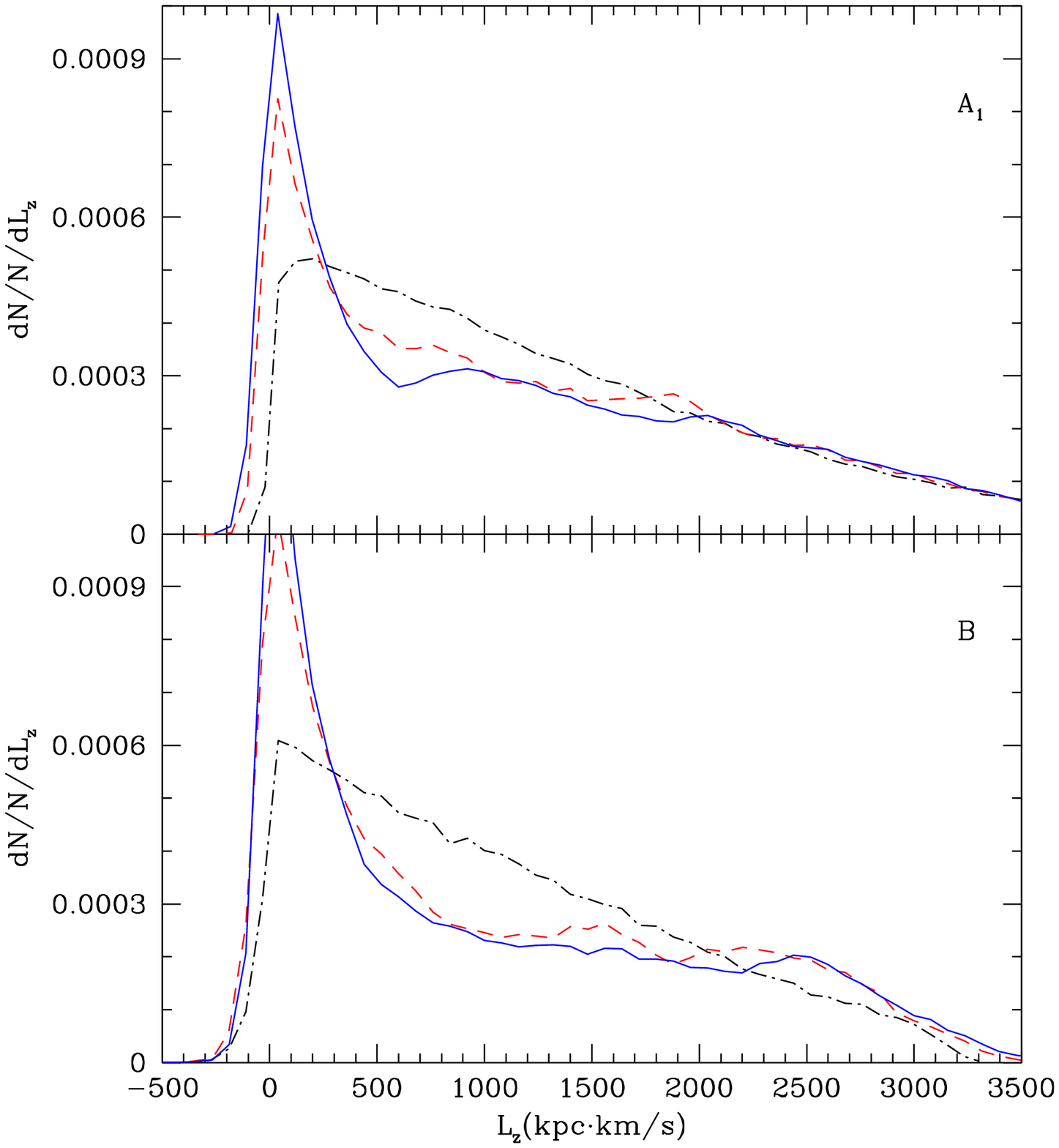}
	\caption{Distribution of the z-component of the angular
	momentum of the stellar particles for models \Aone\ (top
	panel) and \B\ (bottom panel) at different moments of
	time. Initial distribution is shown by dot-dashed
	curves. Particles with low angular momentum are preferentially
	at small radii. Large angular momentum is coming from
	particles at large distances. Dashed curves are for 1-1.5~Gyrs
	after the beginning of the evolution. The final distribution
	(full curves, 4.5-5~Gyrs) is qualitatively the same for both
	models. The peak at small angular momenta corresponds to the
	bar. The number of particles with intermediate angular momenta
	substantially decreases during the evolution. There is an
	excess of particles with very large angular momentum. The
	changes in the distribution are much stronger in the case of
	model \B, which has less dark matter and has a more massive
	disk. } \label{fig:Angmom}
\end{figure}

It was interesting to find that the specific angular momentum $L_z$ at
different radii in Figure~\ref{fig:AngProfile} shows very little
evolution. Only at the very centre the angular momentum declines
slightly.  At larger radii the specific angular momentum is
practically constant. At first sight this indicates that the
distribution of the angular momentum does not evolve. This is not
true. Figure~\ref{fig:Massevol} shows that the stellar mass in the
central region increases quite substantially. Because the specific
angular momentum $L_z(r) $ {\it at a given radius} does not change,
this mass increase implies that the stellar particles that accumulate
in the central region, actually lose their angular momentum. This is
clarified by Figure~\ref{fig:Angmom} that shows significant changes in
the distribution of the angular momentum of the stellar particles in
the course of the evolution. The changes are especially noticeable at
low $L_z$: a large peak forms.  The peak at very low (almost zero)
angular momentum is made of the particles located in the central
region. The particles initially had
intermediate angular momentum and lost some of it  during
the bar formation. As the result of this migration of particles the
distribution of $L_z$ at later moments has a deep at intermediate
angular momenta. The number of particles at large $L_z$ also increases,
which is more clearly seen in the case of model \B.

Evolution of the $L_z$ distribution is not gradual.  Most of the
changes occur before and during the bar formation. Model \B\ clearly
shows that once the bar has formed (after 1~Gyr of evolution) the
distribution experiences very little evolution. This indicates that
the bar itself does not lead to a significant exchange of the angular
momentum between the particles. Most of the evolution is done not by
the bar, but by the processes that produce the bar. Nevertheless, the
evolution does not stop after the bar forms. It proceeds, but with a
much lower rate. 
  

\begin{figure} 
	\includegraphics[width =0.5\textwidth]{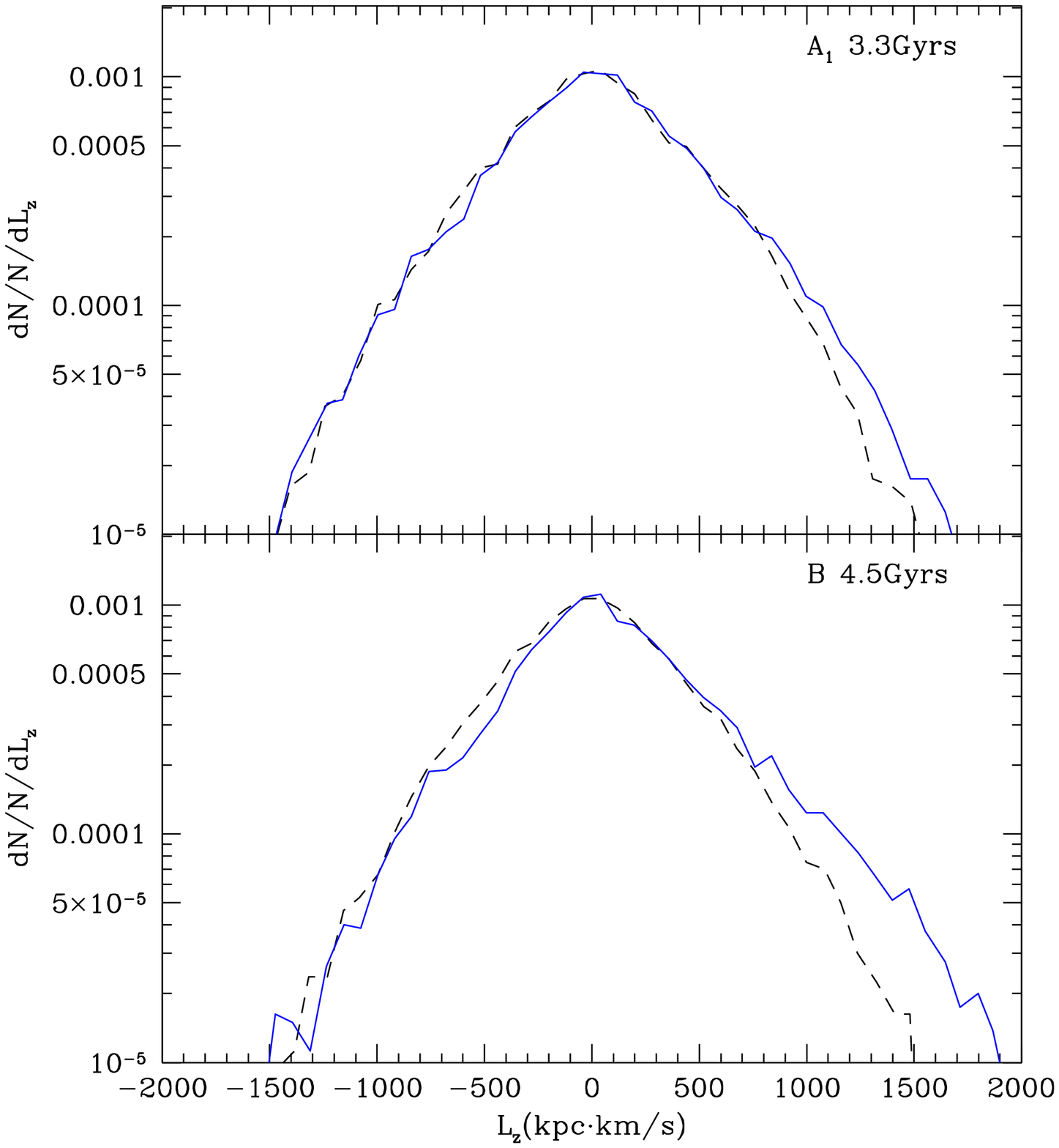} \caption{
	Distribution of z-component of the angular momentum of dark
	matter particles at different moments. Dashed and full curves
	are for initial and advanced moments (3.3 Gyrs for \Aone\ and
	4.5 Gyrs for \B) of evolution. We present
	results only for 10000 particles, which initially were in a
	spherical shell centered at 3~kpc. Particles at other radii
	showed even smaller effects. The changes in the angular
	momentum are clearly seen in both models: there are more
	particles with high angular momentum. This indicates that the
	dynamical friction transfers some angular momentum to the dark
	matter, but the effect is extremely weak. } 
	\label{fig:dmrotation}
\end{figure}

Because of the large moment of inertia of the dark matter, it is
difficult to detect small changes in the distribution of the angular
momentum of the dark matter particles. The largest effect was found in
the central region. By the end of the evolution the dark matter in the
central 2~kpc region rotated with velocity 3 -- 7~$\kms$. This rotation
has very little dynamical effect. Figure~\ref{fig:dmrotation} presents
an example of the change of the angular momentum of the dark matter
particles. It shows the distribution of the $z-$component of
the specific angular momentum of 10,000 particles, which initially
were in a shell centered at 3~kpc from the centre. The change in the
angular momentum is clearly detected, but it is very small.

To summarize, we clearly find indications of the dynamical friction
between the dark matter and the stellar component. Nevertheless, the amount
of the angular momentum lost by the stellar particles is very small:
about 5\% during 5~Gyrs of evolution.  There is much larger exchange
of the angular momentum between stellar particles. Most of this
exchange happens during the formation of the bar. Once the bar is in
place, the distribution of the angular momentum changes very little.

\section{Structure of bars: the pattern speed and the surface density}
\label{sec:BarStructure}
Figure~\ref{fig:Levol} shows the evolution of the pattern speed and
the amplitude of the bars in our simulations. Models \Aone\ and \B\
evolve qualitatively similar. After different periods of time 
models \Aone\ and \B form a bar  which rotates with
almost constant speed for about  5~Gyrs. The amplitude of the bar in  
the model \B\ experienced a noticeable decline (a factor of two) at
2~Gyrs. This change in the amplitude was not accompanied by any other
obvious changes. For example, the pattern speed did not change.

The bar pattern speed is slightly larger in model \B\ ($\approx
33\kms/\kpc$) as compared with model \Aone\ ($\approx
27\kms/\kpc$). What is remarkable about the pattern speeds is that
they do not change much. For example, in model \Aone\ the bar changed
its frequency from $\approx 27\kms/\kpc$ when it formed to $\approx
20\kms/\kpc$ at $t=8.5$~Gyrs.

The evolution of model \Atwo\ is very different. For the first 3~Gyrs
the bar was very quickly rotating with the pattern speed of $\approx
50\kms/kpc$ - almost twice larger than in model \Aone. At later
moments the pattern speed was declining. It went down to $\approx
25\kms/kpc$ by the end of evolution.

\begin{figure} 
\includegraphics[width =0.45\textwidth]{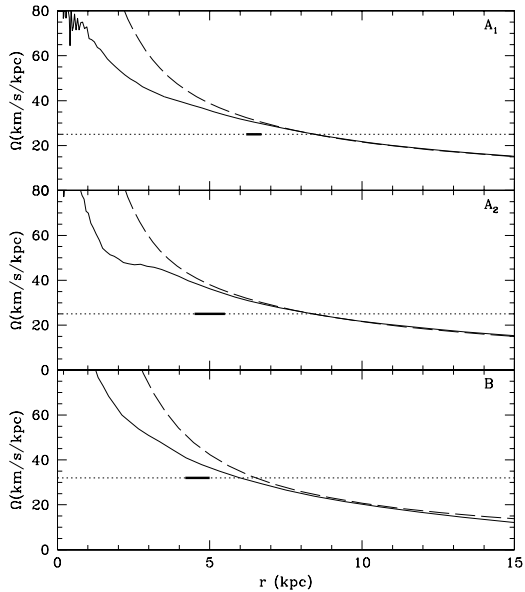}
   \caption{Different frequencies in models \Aone, \Atwo, and \B.  The
   full curves show the frequency of rotation of the disk $\Omega_{\rm
   rot} \equiv V_{\phi}/R$. The dashed curves present the frequency of
   rotation for a cold disk $\Omega_{\rm circ}\equiv \sqrt{GM/R^3}$. The
   difference between the two frequencies are due to the asymmetric
   drift. The dotted horizontal lines show the frequency of rotation of
   the bars.  The extension of the bars in the models is indicated by the
   thick lines.  The distance of a bar from the rotation curves is a
   measure of how fast is the bar. The bar in the model \B\ is clearly a
   very fast bar, which extends almost up to the corotation. The bars in
   models $A$ rotate slower, but they are still fast bars. }
	\label{fig:OmegRot}
\end{figure}

In order to find how fast is the bar rotation, we show in the
Figure~\ref{fig:OmegRot} the bar pattern speeds, the rotation,
and the circular velocity curves for the models. The thick horizontal
bars in the plots indicate the length of the bars. The corotation
radius can be identified as the radius of intersection of the bar
pattern speed with the rotation curve. In all three models the bars
rotate fast with the ratio of the corotation radius to the bar length
equal to $\R \approx 1.2-1.7$.

\begin{figure} 
	\includegraphics[width =0.45\textwidth]{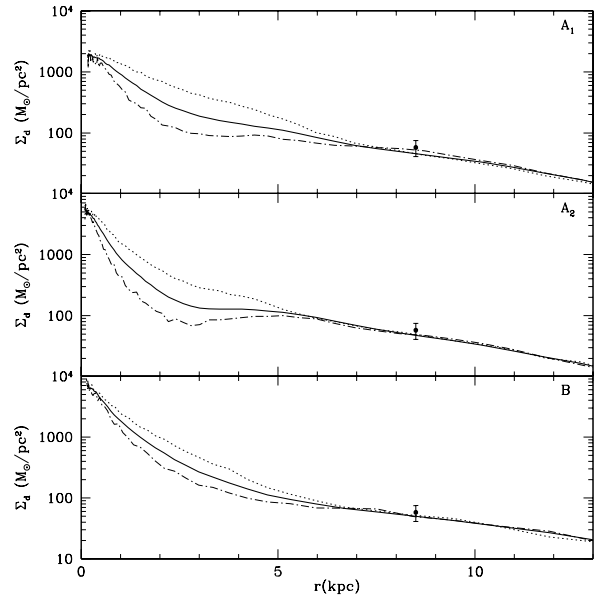}
	\caption{Surface density profiles of the stellar components
	for models \Aone, \Atwo, and \B. The full curves present the
	azimuthally averaged profiles. Dashed curves show double
	exponential fits with parameters presented in the plots. The
	dotted curves show the surface density along the major axes of
	the bars. The surface densities along the minor bar axis are
	shown with the dot-dashed curves. All the curves are obtained
	by averaging over three time moments covering about
	0.5~Gyr. For comparison a large dot shows the surface density
	of the disk (stars + gas) for our Galaxy at the solar
	position. } \label{fig:bulge}
\end{figure}

Figure~\ref{fig:bulge} shows the surface density profiles of the
stellar components in the models. The surface density profiles seems
to indicate a development of normal galactic components: an
exponential disk and a (exponential) bulge. Note that even after
significant evolution in the central part, the outer part of the disk
(radii larger than 5~kpc) is still well approximated by a simple
exponential profile. Still, there is a change even in the outer part:
the exponential scale-length increases quite substantially - by a
factor of 1.5.

\section{Discussion} 
\label{sec:discussion} 

\begin{figure} 
\includegraphics[width =0.5\textwidth]{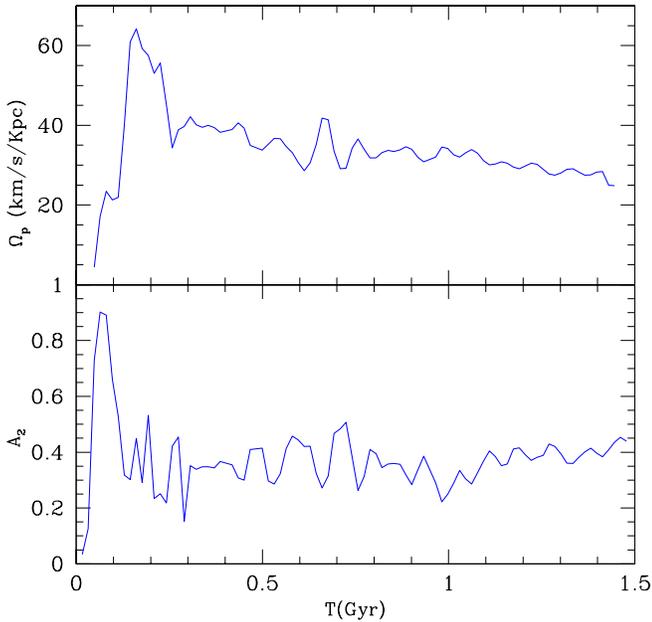}
\caption{Evolution of the pattern speed and the bar amplitude in a
	 simulation with initially cold disk with $Q=0.05$. Other
	 parameters of the simulation are the same as for model
	 \Aone. The simulation was done with a very low resolution of
	 0.5~kpc to mimic results of Debatista \& Sellwood (2000).  In
	 this case the pattern speed quickly decreases with time
	 giving an impression that the bar is slowed down by the
	 dynamical friction with the dark matter. This ``slowing
	 down'' of the bar is an artifact of unrealistically cold disk
	 and grossly insufficient force
	 resolution.}\label{fig:OmegapLR}
\end{figure}

Bars in our models are different from what was found in simulations of
\citet{DB:2000}.  In two of our dark matter-dominated models bars do
not show a strong tendency to slow down. This contradicts results of
\citet{DB:2000}, who always find that the pattern speed of bars
substantially declines with time. We tried to find an explanation for
the disagreement. Numerical effects are the obvious first
suspects. The force resolution in the \citet{DB:2000} models was more 
than an order of magnitude lower than in our simulation. In an effort
to reproduce at least some conclusions of \citet{DB:2000}, we re-run
model \Aone\ with the resolution comparable with that in models of
\citet{DB:2000}. The evolution was so much affected by the lack of the
resolution that after running the model for more than 2~Gyrs and not
finding even a hint of a bar, we abandoned the simulation. We then
re-started the simulation with decreased ``thermal'' velocities in the
disk. The $Q$ parameter was set at the value $Q=0.05$ used by
\citet{DB:2000}. In this simulation a bar
forms. Figure~\ref{fig:OmegapLR} shows that the pattern speed
decreases very substantially over a relatively short period of time -
more or less in line with what was found by \citet{DB:2000}.  Thus, we
find the same behavior of bars -- declining pattern speed -- if we
substantially reduce the resolution and reduce the random velocities
of disk particles. The angular momentum of the stellar particles also  
was declining faster than in the high resolution run. This exercise
clearly demonstrates that the force resolution and the stellar
velocities are  important
factors. In our simulations the low force resolution produced erroneous
results that the bar pattern speed declines with time. The
``coldness'' of the disk may also have affected the pattern speed.
This has also been reported by Athanassoula (2003).  

\begin{figure} 
\includegraphics[width =0.5\textwidth]{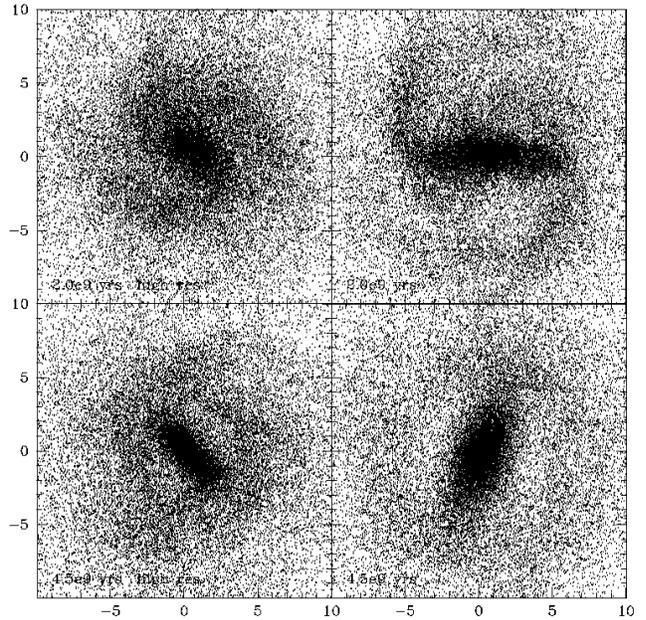}
   \caption{Effects of resolution on bar formation in model \Atwo. Left
   panels show stellar particles in the simulation, which reached 20~pc
   resolution. Right panels show the same model simulated with the
   resolution limited to 350~pc. The top panels are for 2~Gyrs after
   initial moment; the bottom panels are for 4.5~Gyrs.  The high
   resolution run has a very short (1~kpc) bar for 3~Gyrs, while the low
   resolution run has remarkably strong and long bar from the very
   beginning. Axes are labeled in units of kpc. This figure illustrates
   that numerical resolution can significantly affect structure of a bar.
   Lack of resolution results in longer and stronger bars.
}\label{fig:pointsRes}
\end{figure}

Length is another important property of bars. In our simulations bars
have radii comparable to both initial and final exponential disk
length.  This is quite different from what, for example,  
\citet{AthanassoulaMisiriotis:02} find. The bar length in their models
is more than three times longer than the initial disk
scale-length. After 8.5~Gyrs of evolution ($t\approx 610$ in units
used by \citet{AthanassoulaMisiriotis:02}) the bar length in our
\Aone\ model is $6-6.5\kpc$ as compared with $10-14\kpc$ for model MH
in \citet{AthanassoulaMisiriotis:02}.  Bars are also very long in the
models of \citet{DB:2000}: they span the entire disk. Even that a comparison 
with disk scale-length after bar formation could be more favored for
those models it is interesting to test how numerical resolution may
affect the bar length. We rerun model \Atwo\ with a formal resolution
of 350 pc which is 1.6 times the formal resolution of
\citet{AthanassoulaMisiriotis:02}.  Figure~\ref{fig:pointsRes} shows
the distribution of stellar particles of the simulations for model
\Atwo\ at different moments of time. The system was clearly affected 
by the resolution. At 2~Gyrs the bar is barely visible in the high
resolution run, but it is much longer and stronger in the low
resolution run.  The differences are not as large at later moments,
but even at later moments the differences are substantial. This
example shows that the bar length is quite sensitive to the effects of
the resolution.  Short bars are also formed in high resolution
simulations of \citet{Sellwood:02}.

We may only speculate why a better force resolution leads to shorter
and weaker bars. Orbital resonances are traditionally blamed for the
growth of bars. Fast rotating bars trap particles. This in turn
increases the strength of the bar. It gets longer and more massive
trapping even more trajectories.  Low force and high mass resolutions
make the system more resonant by producing a gravitational potential
with too shallow gradients. Motion of particles in such potentials can
be more susceptible to resonances. Thus, we get longer bars.
Increased force resolution results in steeper gradients of the density
and the potential. As the result, particle trajectories do not stay as 
much in resonances, which decreases the rate of the growth of the 
bar. A central concentration makes bars weaker and can even destroy 
them \citep[e.g.,][]{Norman1996}.  During first stages of bar
formation the central concentration increases very dramatically (see
Figures 8 and 9).  A low force resolution artificially makes the
central concentration shallower. This may result in longer and
stronger bars, which are known to rotate slower.  The comparison of
the low and high resolution runs of the model \Atwo\ supports this
picture.

Numerical effects are clearly important for the structure of the  
system. Still, they can not explain all the differences between our 
results and those of \citet{AthanassoulaMisiriotis:02}. The bar in the  
low resolution run of \Atwo\ model is longer than in the high  
resolution run. Still, it is shorter than what was found by 
\citet{AthanassoulaMisiriotis:02}. The remaining differences are likely  
due to the differences in the overall structure of the system. Our 
halo was significantly more extended (250~kpc as compared with 50~kpc 
in \citet{AthanassoulaMisiriotis:02}). Thus, the dark matter 
particles have larger velocities.   
Our disk is also slightly ``hotter''.     
Each effect points in  the same direction: systems studied by  
\citet{AthanassoulaMisiriotis:02} are likely more susceptible to  
resonances. The effects accumulate over billions of years producing    
longer bars in \citet{AthanassoulaMisiriotis:02}.    
\citet{ Athanassoula2003} presents a similar and more extensive discussion      
which conclude  that  a hotter system is  less resonant  and 
has a slower decline of the  bar pattern speed  and a lower rate of bar growth.    
In particular \citet{ Athanassoula2003} shows that  bars in disks with a   
Toomre parameter  $Q =.1$ slow down more thant four times faster than in a  
disk with  $Q =2.2$, this result is in agreement with our discussion.    
We can conclude that the structure  and kinematics of  halo  
and disk have a central role in bar evolution. Differences within our simulations   
and previous works like  \citet{AthanassoulaMisiriotis:02}) are influenced not  
only by resolution but also by the different structure of the models.      

It is still  a matter of debate which initial setup is more realistic. Our  
halos are consistent with predictions of cosmological models. They   
extend to the virial radius; they have realistic profiles and rotation curves. 
Halos in \citet{AthanassoulaMisiriotis:02} produce reasonable rotation curves 
although they are not consistent with cosmological models.  
In any case, we do not confirm  
\citet{AthanassoulaMisiriotis:02} conclusions that bars in models with   
centrally concentrated dark matter are necessarily too long (3-4 disk  
initial exponential lengths). Their results are valid only for the    
particular models they studied. They are not valid as a general   
statement that realistic dark matter models should have excessively  
long bars.  While we have mentioned some disagreements with  
\citet{AthanassoulaMisiriotis:02}, it should be emphasized, that we  
find many similar results at least on the qualitative level. In 
agreement with \citet{DB:2000} and \citet{AthanassoulaMisiriotis:02} 
we find that a massive dark matter halo does not prevent formation of 
a strong bar. We also find that bars have a tendency to grow with time   
and do not dissolve spontaneously.  We agree with  
\citet{AthanassoulaMisiriotis:02} that formation of a bar leads to a 
baryon-dominated central region even in cases when the disk is  
initially submaximal. We also find that formation and evolution of 
bars does not lead to a decline in the density of the dark matter in 
the central region.   

It is an interesting question how a bar can possibly {\it not} slow 
down when it rotates inside a dense dark matter halo. Even in 
model \Aone\ the disk lost 6 percent of its angular momentum 
after 5~Gyrs of evolution. One should expect that most of the loss 
comes from the bar. Because the bar has only 40\% of the stellar mass, 
one expects that the bar should lose  around 15\% of its 
angular momentum. This is still marginally consistent with the
behavior of the pattern speed in Figure~\ref{fig:Levol}. Still, what
if $\Omega_p$ is really constant? Can a bar rotate with the constant
speed and lose its angular momentum? If we believe the arguments
presented by \citet{Weinberg:85}, the answer is ``No''. Indeed, if a
solid rotating bar, often used in these kind of estimates
\citep{Weinberg:85, LittleCarlberga:91,WeinbergKatz:2001}, loses its
angular momentum, it {\sl must} slow down. Yet, real bars may behave
differently because they are not solid objects. Bars are made of
particles, which stream through them creating a wave pattern, which we
call a ``bar''. When the bar loses angular momentum, it is individual
particles which lose it. The particles can move to orbits with smaller
radii where orbital frequency is larger. It is reasonable to assume
that the bar speed is proportional to the orbital frequency of
particles. Because the particles rotate with larger frequency, the bar
can speed {\it up} when it loses the angular momentum. Real bars have
another feature, which complicates things even more. Typically, they
have a tendency to grow. For example, the bar in model \Aone\ has
increased its length from $\approx 4-5\kpc$ at 3~Gyrs to $\approx
6-6.5\kpc$ at 5~Gyrs. This 30\% increase in the bar length is due to
particles, that came from large radii. Because the rotation curve is
nearly flat, particles at larger radii have smaller orbital
frequencies. When they ``join'' the bar, it will have a tendency to
slow down. This happens if the particles do not significantly change
trajectories. If the particles lose their angular momentum and
decrease average radius, they may cause the bar to speed up.

The outcome of all these competing processes is difficult to
predict. We find that slow growing bars rotate with almost constant
speed (e.g., bars in models \Aone, \B, and the bar in model \Atwo\ at
$t<3$~Gyrs). If a bar grows fast, it significantly decreases its
pattern speed. This happened in model \Atwo\ at $t=3-5$~Gyrs. At 3~Gyr
the bar was very short ($\sim 1.5\kpc$) and weak. At earlier moments
of time it was growing very slowly and was rotating with high constant
speed for almost 3~Gyrs. At $t\approx 3$~Gyrs the bar starts to grow
very fast. By the end of evolution (6~Gyrs) it was 4.5 -- 5\kpc\ long
and it slowed down by a factor of two. 

Do bars in our models rotate as fast as the observed bars? In our
models bars have $\R =1.2 - 1.7$. This should be compared with the
observational results. Unfortunately the number of galaxies, for which
the direct method of \citet{TremainWeinberg:84} was used, is small and
the errors are still large. For NGC 936 (Sa type)
\citet{Merrifield1995} find $\R =1.4\pm 0.3$. \citet{Gerssen1999} find
$\R =1.15^{+0.38}_{-0.23}$ for S0 galaxy NGC 4596. For five S0
galaxies \citet{Aguerri2003} find that ``$\R$~ is consistent with being
in the range from 1.0 to 1.4''. Taken at the face value, these results
are not consistent with most of our bars, which typically have $\R
\approx 1.5$. Yet, on pure statistical grounds our bars are still
acceptable because the errors of the quoted observational results are
only $1\sigma$'s. For example, at the $2\sigma$ level, the five
galaxies studied by \citet[][Table8]{Aguerri2003} give the following
upper limits: $\R<1.4$ (ESO 139-G009), $<2.8$ (IC 874), $<1.6$ (NGC
1308),  $<1.9$ (NGC 1440), and $<2.3$ (NGC 3412). Our bars are well within the limits.

There is another alarming issue: the overall structure of early type
-- S0 and Sa -- spirals studied in the observations is very different
from that of the numerical models. The later aim at Sb and Sc types,
not S0.  The surface density in simulations is well described by the
double exponential law (Figure~15) ; the stellar velocity dispersion
is relatively small (Figures~2-3). The situation is very different
for the galaxies studied by \citet{Aguerri2003}. Surface brightness
profiles are not even close to double exponentials. Every galaxy
requires four components: a central bulge, a bar, a lens, and a disk
\citep[Figure 7,][]{Aguerri2003}. For example, NGC 3412, which has
circular velocity $205\kms$ (close to our models), is dominated in the
central $\sim 1\kpc$ region by a bulge, which does not have any
counterpart in the simulations. A massive lens component is important
at around $3\kpc$. Again it does not exist in our models.
Line-of-sight velocities are also significantly larger than what we have in
the models. While it is clear that we are dealing with very different
objects, one may only speculate how these differences affect $\R$.
\citet{CombesElmegreen:93} find that early type spirals should have
faster rotating bars. If that is confirmed for models with live bulges
and halos, disagreements with observations may become smaller. In the
same direction point results of \citet{Athanassoula2003}, who finds
that bars in hotter disks rotate faster. 

In order to resolve the issues, numerical models should be improved to
 mimic S0 galaxies. Models likely should include a small initial bulge
 and should have disks with larger random velocities. Observations
 should also be done for Sb and Sc galaxies. Before this is done, it
 is difficult to assess the situation. As it stands now, differences
 with observations are not large and models cannot be rejected.

\section{Conclusions}
\label{sec:conclusions}

We study formation of bars in galactic models with equilibrium
exponential disks rotating inside extended dark matter halos predicted
by modern cosmological models. The amount of the dark matter in the
central 10~kpc region is substantial - it is larger than the disk
mass. Our models have parameters, which approximately match those of
the Milky Way galaxy: the virial mass is $(1-2)\times 10^{12}\Msun$,
the total mass of the disk and the bulge is $(4-6) \times
10^{12}\Msun$. Dark matter halos in our models extend to the virial
radius of 200-300\kpc and have cuspy inner profiles. We use millions
of particles to simulate the models.

Extended strong bars form in all our models in spite of the fact that 
the random velocities are much larger than the kinetic energy of disk  
rotation. Our results are at odds with the Ostriker-Peebles criterion  
of bar stability.
We find that the dynamical friction between the  
dark matter and the bar results in some transfer of angular momentum
to the dark matter halo, but the effect is much smaller then
previously found in low resolution simulations and is incompatible
with early analytical estimates. The mass and the force resolutions
are crucial for the dynamics of bars. In simulations with the low
resolution of 300-500~pc we find that the bar slows down and the
angular momentum is lost relatively fast. In simulations with millions
of particles, which reach the resolution of 20-40~pc, the pattern
speed may not change over billions of years and the stellar component
loses very little (5 -- 10\%) of its total angular momentum.

Bars in our models are fast rotators with the ratio of the corotation
radius to the bar major semi-axis being in the range $\R =1.2-1.7$,
which is marginally compatible with observational results.
The initial random velocities of the disk in the central $\approx
1-2\kpc$ region can substantially affect the structure of bars. In our
simulations ``hot'' disks develop a longer bar, which rotates at an
almost constant pattern speed. Colder models generate an initially
shorter bar, which grows and slows down.

In contrast to many previous simulations, bars in our models are
relatively short. As in many observed cases, the major bar semi-axis
in the models is about an exponential length of the disk.

The transfer of the angular momentum between inner and outer parts of
the disk plays a very important role in the secular evolution of 
disk. The main effect is the angular momentum transfer from the
central part of the bar to the outer disk. This transfer leads to a
substantial increase of the stellar mass and to a decrease of the dark
matter-to-baryons ratio in the centre of the galaxy. For the Milky
Way- scale models the central 2~kpc is strongly dominated by the
baryonic component after 1 -- 2~Gyrs since the onset of the
instability. At intermediate 3 --10~kpc scales the disk is
sub-dominant: spherically average density of the disk is 2-3 times
smaller than the density of the dark matter. This makes an interesting
twist for the never ending debate between supporters and opponents of
the maximum disks. We predict that  barred galaxies (or galaxies, which were
barred) in the central region ( $\sim$~1/2 of the disk exponential
length) should have the maximum disk (maximum bulge or bar, to be more
precise), but the outer part is the sub-maximum disk. 

 Another effect of the bar formation is the increase by a factor of
$\approx 1.2-1.5$ of the exponential length of the disk. This increase
may change theoretical predictions for the disk lengths in
hierarchical models.

We find that the surface density profile of an evolved system is well
approximated by a double exponential law. The 1/4 law for the bulge
gives a worse fit, but it is not excluded. Only in extreme cases (very
late stages of evolution, very strong bars formed in low resolution
simulations) we see a tendency for a bar to produce a flat part of the
surface density in the interface of the bar and the disk.

To summarize, we use more realistic models than in most of previous
simulations and find that the models with substantial amount of the
dark matter produce bar structure in striking agreement with
observational results.

\section*{Acknowledgments}

We thank J. Sellwood for exciting discussions and for showing us his 
preliminary results. We thank L. Athanassoula, Stephane Courteau
and Bruce Elmegreen for numerous comments and suggestions for improving the draft of our
paper. We are grateful to N. Vogt for her comments. We thank the Center for 
Cosmological Physics, University of Chicago, for hospitality and
support.  We acknowledge support from the grants NAG- 5- 3842 and NST-
9802787 to NMSU.  Computer simulations presented in this paper were
done at the National Center for Supercomputing Applications (NCSA) at
Urbana-Champaign and at the National Energy Research Scientific
Computing Center at the Lawrence Berkeley National Laboratory.

\bibliographystyle{mn2e}
\bibliography{bars}

\end{document}